\documentstyle{siam} 

\newcommand{\rake}[0]{{\cal R}}
\newcommand{\ltube}[0]{{\cal L}}
\newcommand{\remark}[0]{{\it Note}}
\newcommand{\case}[0]{{\it Case}}

\newcommand{\Ab}[0]{\overline{A}}
\newcommand{\uhalf}[2]{#1,\frac{#1+#2-1}{2}}
\newcommand{\lhalf}[2]{\frac{#1+#2+1}{2},#2}

\newtheorem{fact}{Fact}
\newtheorem{problem}{Problem}

\newcommand{\rr}[0]{\mbox{\sc rr}}
\newcommand{\ra}[0]{\mbox{\sc ra}}
\newcommand{\rp}[0]{\mbox{\sc rp}}
\newcommand{\PP}[0]{{\cal P}}
\newcommand{\QQ}[0]{{\cal Q}}

\newcommand{\mm}[0]{\mbox{\sc m}}
\newcommand{\mmb}[0]{\overline{\mbox{\sc m}}}

\newcommand{\ALGORITHM}[0]{{\bf Algorithm}}
\newcommand{\INPUT}[0]{{\bf Input}}
\newcommand{\OUTPUT}[0]{{\bf Output}}
\newcommand{\BEGIN}[0]{{\bf begin}}
\newcommand{\END}[0]{{\bf end}}

\newcommand{\WHILE}[0]{{\bf while}}

\newcommand{\DO}[0]{{\bf do}}

\begin{document}

\title{Tree Contractions and Evolutionary Trees}

\author{Ming-Yang Kao\thanks{Department of Computer Science, Yale
University, New Haven, CT 06520 (kao-ming-yang@cs.yale.edu). This
research was supported in part by NSF grant CCR-9531028.}}

\maketitle

\pagestyle{myheadings}
\markboth{\sc m.~y.~kao}{\sc tree contractions and evolutionary trees}

\begin{abstract}
  An {\it evolutionary tree} is a rooted tree where each internal vertex
  has at least two children and where the leaves are labeled with distinct
  symbols representing species.  Evolutionary trees are useful for modeling
  the evolutionary history of species.  An {\it agreement subtree} of two
  evolutionary trees is an evolutionary tree which is also a topological
  subtree of the two given trees.  We give an algorithm to determine the
  largest possible number of leaves in any agreement subtree of two trees
  $T_1$ and $T_2$ with $n$ leaves each.  If the maximum degree $d$ of these
  trees is bounded by a constant, the time complexity is $O(n\log^2{n})$
  and is within a $\log{n}$ factor of optimal. For general $d$, this
  algorithm runs in $O(nd^2\log{d}\log^2{n})$ time or alternatively in
  $O(nd\sqrt{d}\log^3{n})$ time.
\end{abstract}


\begin{AMSMOS} 
05C05, 05C85, 05C90, 68Q25, 92B05
\end{AMSMOS}

\section{Introduction}
An {\it evolutionary tree} is a rooted tree where each internal vertex has
at least two children and where the leaves are labeled with distinct
symbols representing species.  Evolutionary trees are useful for modeling
the evolutionary history of species.  Many mathematical biologists and
computer scientists have been investigating how to construct and compare
evolutionary trees \cite{ AF94, ASSU81, BFW92, DayS87, Dress92, FKW95,
  Felsenstein82, Felsenstein83, Felsenstein88, Friday89, Gusfield91,
  Hein89, Hendy89, HP82, JLW94, KLW90, KW94, KA94, KB81, PH86, RM92,
  Steel92, WJL96, Warnow94}.  An {\it agreement subtree} of two evolutionary trees
is an evolutionary tree which is also a topological subtree of the two
given trees.  A {\it maximum} agreement subtree is one with the largest
possible number of leaves.  Different theories about the evolutionary
history of the same species often result in different evolutionary trees.
A fundamental problem in computational biology is to determine how much two
theories have in common.  To a certain extent, this problem can be answered
by computing a maximum agreement subtree of two given evolutionary trees
\cite{FG85}.

Let $T_1$ and $T_2$ be two evolutionary trees with $n$ leaves each.  Let
$d$ be the maximum degree of these trees.  Previously, Kubicka, Kubicki and
McMorris \cite{KKM94} gave an algorithm that can compute the number of
leaves in a maximum agreement subtree of $T_1$ and $T_2$ in
$O(n^{(\frac{1}{2}+\epsilon)\log{n}})$ time for $d=2$.  Steel and Warnow
\cite{SW93} gave the first polynomial-time algorithm.  Their algorithm runs
in $O(\min\{d!n^2,d^{2.5}n^2\log n\})$ time if $d$ is bounded by a constant
and in $O(n^{4.5}\log{n})$ time for general trees. Farach and Thorup
\cite{FT94a} later reduced the time complexity of this algorithm to
$O(n^2)$ for general trees. More recently, they gave an algorithm
{\cite{FT94b}} that runs in $O(n^{1.5}\log n)$ time for general trees.  If
$d$ is bounded by a constant, this algorithm runs in $O(n c^{\sqrt{\log
    n}}+n\sqrt{d}\log n)$ time for some constant $c > 1$.

This paper presents an algorithm for computing a maximum agreement subtree
in $O(n\log^2{n})$ time for $d$ bounded by a constant.  Since there is a
lower bound of $\Omega(n\log{n})$, our algorithm is within a $\log{n}$
factor of optimal. For general $d$, this algorithm runs in
$O(nd^2\log{d}\log^2{n})$ time or alternatively in $O(nd\sqrt{d}\log^3{n})$
time.  This algorithm employs new tree contraction techniques \cite{ADKP89,
  GMT88, KD88, MR89, MR91}.  With tree contraction, we can immediately
obtain an $O(n\log^5{n})$-time algorithm for $d$ bounded by a constant.
Reducing the time bound to $O(n \log^2n)$ requires additional techniques.
We develop new results that are useful for bounding the time complexity of
tree contraction algorithms.  As in \cite{FT94a, FT94b, SW93}, we also
explore the dynamic programming structure of the problem.  We obtain some
highly regular structural properties and combine these properties with the
tree contraction techniques to reduce the time bound by a factor of
$\log^2n$. To remove the last $\log n$ factor, we incorporate some
techniques that can compute maxima of multiple sets of sequences at
multiple points, where the input sequences are in a compressed format.

We present tree contraction techniques in \S\ref{sec_rake} and outline our
algorithms in \S\ref{sec_tree_def}.  The maximum agreement subtree problem
is solved in \S\ref{sec_reductions} and \S\ref{sec_one_one} with a
discussion of condensed sequence techniques in \S\ref{sec_condensed}.
Section \S\ref{sec_disc} concludes this paper with an open problem.

\section{New tree contraction techniques}\label{sec_rake}
Throughout this paper, all trees are rooted ones, and every nonempty tree
path is a vertex-simple one from a vertex to a descendant.  
For a tree $T$ and a vertex $u$, let $T^u$ denote the subtree of $T$ formed
by $u$ and all its descendants in $T$.  

A key idea of our dynamic programming approach is to partition $T_1$ and
$T_2$ into well-structured tree paths. We recursively solve our problem for
$T_1^x$ and $T_2^y$ for all heads $x$ and $y$ of the tree paths in the
partitions of $T_1$ and $T_2$, respectively.  The partitioning is based on
new tree contraction techniques developed in this section.

\begin{figure}
\begin{center}\vspace{5.00cm}

\begin{picture}(50,40)(0,0)
\setlength{\unitlength}{.2cm}
\put(-20,05){\circle*{2}}\put(-18,05){$x_1$}
\put(-15,15){\circle*{2}}\put(-13,15){$x_2$}
\put(-10,05){\circle*{2}}\put(-08,05){$x_3$}
\put(-05,25){\circle*{2}}\put(-03,25){$x_4$}
\put(-00,10){\circle*{2}}\put( 02,10){$x_5$}
\put(005,15){\circle*{2}}\put(007,15){$x_6$}
\put(005,30){\circle*{2}}\put(007,30){$x_7$}
\put(010,10){\circle*{2}}\put(012,10){$x_8$}
\put(010,25){\circle*{2}}\put(012,25){$x_9$}
\put(015,35){\circle*{2}}\put(017,35){$x_{10}$}
\put(020,20){\circle*{2}}\put(022,20){$x_{11}$}
\put(020,30){\circle*{2}}\put(022,30){$x_{12}$}

\put(-15,15){\line(-1,-2){ 5}}
\put(-15,15){\line( 1,-2){ 5}}

\put(-05,25){\line(-1,-1){10}}
\put(-05,25){\line( 1,-1){10}}

\put(005,15){\line(-1,-1){ 5}}
\put(005,15){\line( 1,-1){ 5}}

\put(005,30){\line(-2,-1){10}}
\put(005,30){\line( 1,-1){ 5}}

\put(015,35){\line(-2,-1){10}}
\put(015,35){\line( 1,-1){ 5}}

\put(020,20){\line( 0, 1){10}}
\end{picture}

After the first rake, the above tree becomes the following tree.
\end{center}

\begin{center}\vspace{5.00cm}

\begin{picture}(50,30)(0,0)
\setlength{\unitlength}{.2cm}
\put(-15,05){\circle*{2}}\put(-13,05){$x_2$}
\put(-05,15){\circle*{2}}\put(-03,15){$x_4$}
\put(005,05){\circle*{2}}\put(007,05){$x_6$}
\put(005,20){\circle*{2}}\put(007,20){$x_7$}
\put(015,25){\circle*{2}}\put(017,25){$x_{10}$}

\put(-05,15){\line(-1,-1){10}}
\put(-05,15){\line( 1,-1){10}}

\put(005,20){\line(-2,-1){10}}

\put(015,25){\line(-2,-1){10}}
\end{picture}

After the second rake, the above tree becomes the following tree.
\end{center}

\begin{center}\vspace{3.50cm}

\begin{picture}(50,20)(0,0)
\setlength{\unitlength}{.2cm}
\put(-05,05){\circle*{2}}\put(-03,05){$x_4$}
\put(005,10){\circle*{2}}\put(007,10){$x_7$}
\put(015,15){\circle*{2}}\put(017,15){$x_{10}$}

\put(005,10){\line(-2,-1){10}}

\put(015,15){\line(-2,-1){10}}
\end{picture}

After the third rake, the above tree becomes empty.  

\end{center}
\caption{An example of iterative applications of rakes.}
\label{fig_rake}
\end{figure}
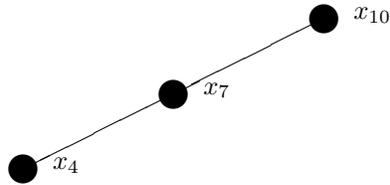

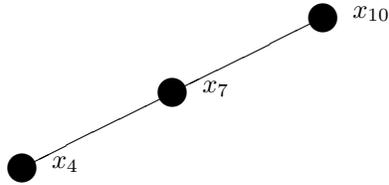
\begin{figure}
\begin{center}\vspace{5.00cm}

\begin{picture}(50,40)(0,0)
\setlength{\unitlength}{.2cm}
\put(-20,05){\circle*{2}}\put(-18,05){$x_1$}
\put(-10,05){\circle*{2}}\put(-08,05){$x_3$}
\put(-00,10){\circle*{2}}\put( 02,10){$x_5$}
\put(010,10){\circle*{2}}\put(012,10){$x_8$}
\put(010,25){\circle*{2}}\put(012,25){$x_9$}
\put(020,20){\circle*{2}}\put(022,20){$x_{11}$}
\put(020,30){\circle*{2}}\put(022,30){$x_{12}$}

\put(020,20){\line( 0, 1){10}}
\end{picture}

The first rake deletes the above leaf tubes.
\end{center}

\begin{center}\vspace{5.00cm}

\begin{picture}(50,30)(0,0)
\setlength{\unitlength}{.2cm}
\put(-15,05){\circle*{2}}\put(-13,05){$x_2$}
\put(005,05){\circle*{2}}\put(007,05){$x_6$}
\end{picture}

The second rake deletes the above leaf tubes. 
\end{center}

\begin{center}\vspace{3.50cm}

\begin{picture}(50,20)(0,0)
\setlength{\unitlength}{.2cm}
\put(-05,05){\circle*{2}}\put(-03,05){$x_4$}
\put(005,10){\circle*{2}}\put(007,10){$x_7$}
\put(015,15){\circle*{2}}\put(017,15){$x_{10}$}

\put(005,10){\line(-2,-1){10}}

\put(015,15){\line(-2,-1){10}}
\end{picture}

The third rake deletes the above leaf tube.

\end{center}
\caption{The leaf tubes deleted by the rakes in
  Figure~\protect{\ref{fig_rake}}.}
\label{fig_tube}
\end{figure}

A tree is {\it homeomorphic} if every internal vertex of that tree has at
least two children.  Note that the size of a homeomorphic tree is less than
twice its number of leaves.  Let $S$ be a tree that may or may not be
homeomorphic. A {\it chain of $S$} is a tree path in $S$ such that every
vertex of the given path has at most one child in $S$.  A {\it tube of $S$}
is a maximal chain of $S$.  A {\it root path of a tree} is a tree path
whose head is the root of that tree; similarly, a {\it leaf path} is one
ending at a leaf.  A {\it leaf tube of $S$} is a tube that is also a leaf
path.  Let $\ltube(S)$ denote the set of leaf tubes in $S$.  Let $\rake(S)
= S-\ltube(S)$, i.e., the subtree of $S$ obtained by deleting from $S$ all
its leaf tubes.  The operation $\rake$ is called the {\it rake operation}.
See Figures~\ref{fig_rake} and \ref{fig_tube} for examples of rakes and
leaf tubes.

Our dynamic programming approach iteratively rakes $T_1$ and $T_2$ until
they become empty.  The tubes obtained in the process form the desired
partitions of $T_1$ and $T_2$.  Our rake-based algorithms focus on certain
sets of tubes described here.  A {\it tube system} of a tree $T$ is a set
of nonempty tree paths $P_1,\cdots,P_m$ in $T$ such that (1) the paths
$P_i$ contain no leaves of $T$ and (2) $T^{h_1},\cdots,T^{h_m}$ are
pairwise disjoint, where $h_i$ is the head of $P_i$.  Condition (1) is
required here because our rake-based algorithms process leaves and non-leaf
vertices differently. Condition (2) holds if and only if for all $i$ and
$j$, $h_i$ is not an ancestor or descendant of $h_j$.  We can iteratively
rake $T$ to obtain tube systems.  The set of tubes obtained by the first
rake, i.e., $\ltube(T)$, is not a tube system of $T$ because $\ltube(T)$
simply consists of the leaves of $T$ and thus violates Condition (1).
Every further rake produces a tube system of $T$ until $T$ is raked to
emtpy.  Our rake-based algorithms only use these systems although there may
be others.

We next develop a theorem to bound the time complexities of rake-based
algorithms in this paper.  For a tree path $P$ in a tree $T$,
\begin{itemize}
\item $K(P,T)$ denotes the set of children of $P$'s vertices in $T$,
  excluding $P$'s vertices;
\item
$t(P)$ denotes the number of vertices in $P$;
\item 
$b(P,T)$ denotes the number of leaves in $T^h$ where $h$ is the head of
$P$.
\end{itemize}
(The symbol $K$ stands for the word kids, $t$ for top, and $b$ for bottom.)

Given $T$, we recursively define a mapping $\Phi_T$ from the subtrees $S$
of $T$ to reals.  If $S$ is an empty tree, then $\Phi_T(S)=0$. Otherwise,
\[\Phi_T(S) = \Phi_T(\rake(S))+\sum_{P \in 
\ltube(S)}b(P,T)\cdot\log(1+t(P)).\] 

({\it Note.} All logarithmic functions $\log$ in this paper are in base
$2$.)

\begin{theorem}\label{thm_rake}
  For all positive integers $n$ and all $n$-leaf homeomorphic trees $T$,
  $\Phi_T(T)\leq n(1+\log{n})$.
\end{theorem}
\begin{proof}
For any given $n$, $\Phi_T(T)$ is maximized when $T$ is a binary tree
formed by attaching $n$ leaves to a path of $n-1$ vertices.  The proof is
by induction.

{\it Base Case.} For $n = 1$, the theorem trivially holds.

Now assume $n \geq 2$.

{\it Induction Hypothesis.} For every positive integer $n' < n$, the
theorem holds.

{\it Induction Step.} Let $r$ be the smallest integer such that $T$ is
empty after $r$ rakes.  Then, at the end of the $(r-1)$-th rake, $T$ is a
path $P=x_1,\cdots,x_p$. Let $T_1,\cdots,T_s$ be the subtrees of $T$ rooted
at vertices in $K(P,T)$.  Let $n_i$ be the number of leaves in $T_i$.  Note
that
\[
\Phi_T(T)=n\log(p+1)+\sum_{i=1}^{s}\Phi_{T_i}(T_i).
\]
Since $1 \leq n_i < n$ and $T_i$ is homeomorphic, by the induction
hypothesis,
\[
\Phi_T(T) \leq n\log(p+1)+
\sum_{i=1}^{s}n_i(1+\log{n_i}).
\]
Since $\sum_{i=1}^{s}n_i = n$,
\begin{eqnarray}\label{eq_rake_2}
\Phi_T(T) \leq n+n\log(p+1)+\sum_{i=1}^{s}n_i\log{n_i}.
\end{eqnarray}
Because $T$ is homeomorphic, each $x_i$ has at least one child in $K(P,T)$.
Since $n \geq 2$, $r \geq 2$. Then, $x_p$ cannot be a leaf in $T$ and thus
has at least two children in $K(P,T)$.  Consequently, $s \geq p+1$.  Next,
note that for all $m_1,m_2 > 0$,
\[
m_1\log{m_1} +m_2\log{m_2} \leq (m_1+m_2)\log(m_1+m_2).
\]
With this inequality and the fact that $s \geq p+1$, we can combine the
terms in the right-hand side summation of Inequality~\ref{eq_rake_2} to
obtain the following inequality.
\begin{eqnarray}\label{eq_rake_3}
\Phi_T(T) \leq n+n\log(p+1)+\sum_{i=1}^{p+1}n'_i\log{n'_i},
\end{eqnarray}
where $\sum_{i=1}^{p+1}n'_i = n$ and $n'_i \geq 1$.  For any given $p$, the
summation in Inequality~\ref{eq_rake_3} is maximized when $n'_1=n-p$ and
$n'_2=\cdots=n'_{p+1}=1$.  Therefore,
\begin{eqnarray}\label{eq_rake_4}
\Phi_T(T) \leq n+n\log(p+1)+(n-p)\log(n-p).
\end{eqnarray}
The right-hand side of Inequality~\ref{eq_rake_4} is maximized when
$p=n-1$. This gives the desired bound and finishes the induction proof.
\end{proof} 

\section{Comparing evolutionary trees}\label{sec_tree_def}
Formally, an {\it evolutionary tree} is a homeomorphic tree whose leaves
are labeled by distinct labels. The {\it label set of an evolutionary tree}
is the set of all the leaf labels of that tree.

The {\it homeomorphic version $T'$ of a tree $T$} is the homeomorphic tree
constructed from $T$ as follows. Let $W = \{w~|~w$ is a leaf of $T$
  or is the lowest common ancestor of two leaves$\}$.  $T'$ is the tree over
$W$ that preserves the ancestor-descendant relationship of $T$.  Let $T_1$
and $T_2$ be two evolutionary trees with label sets $L_1$ and $L_2$,
respectively.
\begin{itemize}
\item For a subset $L_1'$ of $L_1$, $T_1||L_1'$ denotes the homeomorphic
  version of the tree constructed by deleting from $T_1$ all the leaves
  with labels outside $L'_1$.
\item
Let $T_1||T_2 = T_1||(L_1{\cap}L_2)$.
\item
For a tree path $P$ of $T_1$, $P||T_2$ denotes the tree path in $T_1||T_2$
formed by the vertices of $P$ that remain in $T_1||T_2$.
\item 
For a set $\PP$ of tree paths $P_1,\cdots,P_m$ of $T_1$, $\PP||T_2$ denotes
the set of all $P_i||T_2$.
\end{itemize}

Formally, if $L'$ is a maximum cardinality subset of $L_1{\cap}L_2$ such
that there exists a label-preserving tree isomorphism between $T_1||L'$ and
$T_2||L'$, then $T_1||L'$ and $T_2||L'$ are called {\it maximum agreement
  subtrees of $T_1$ and $T_2$}.
\begin{itemize}
\item
$\rr(T_1,T_2)$ denotes the number of leaves in a maximum agreement subtree
of $T_1$ and $T_2$. 
\item $\ra(T_1,T_2)$ is the mapping from each vertex $v \in T_2||T_1$ to
  $\rr(T_1,(T_2||T_1)^v)$, i.e., $\ra(T_1,T_2)(v)=\rr(T_1,(T_2||T_1)^v)$.
\end{itemize}
For a tree path $Q$ of $T_2$, if $Q$ is nonempty, let $H(Q,T_2)$ be the set
of all vertices in $Q$ and those in $K(Q,T_2)$.  If $Q$ is empty, let
$H(Q,T_2)$ consist of the root of $T_2$, and thus, if both $T_2$ and $Q$
are empty, $H(Q,T_2)=\emptyset$.
\begin{itemize}
\item For a set $\QQ$ of tree paths $Q_1,\cdots,Q_m$ of $T_2$, let
  $\rp(T_1,T_2,\QQ)$ be the mapping from $v \in \cup_{i=1}^{m}
  H(Q_i||T_1,T_2||T_1)$ to $\rr(T_1,(T_2||T_1)^v)$, i.e.,
  $\rp(T_1,T_2,\QQ)(v) = \rr(T_1,(T_2||T_1)^v)$.  For simplicity, when
  $\QQ$ consists of only one path $Q$, let $\rp(T_1,T_2,Q)$ denote
  $\rp(T_1,T_2,\QQ)$.
\end{itemize}
(The notations $\rr$, $\ra$ and $\rp$ abbreviate the phrases root to root,
root to all and root to path.  We use $\rr$ to replace the notation ${\sc
  mast}$ of previous work {\cite{FT94a,FT94b,SW93}} for the sake of
notational uniformity.)

\begin{lemma}\label{lem_tree_restrict}
Let $T_1,T_2,T_3$ be evolutionary trees.
\begin{itemize}
\item
$(T_1||T_2)||T_3=T_1||(T_2||T_3).$
\item
If $T_3$ is a subtree of $T_1$, then $T_3||T_1=T_1||T_3=T_3$.
\item
$\rr(T_1,T_2)=\rr(T_1||T_2,T_2)=\rr(T_1,T_2||T_1)=\rr(T_1||T_2,T_2||T_1)$.
\end{itemize}
\end{lemma}
\begin{proof}
Straightforward.
\end{proof}

\begin{fact}[\cite{FT94a}] 
\label{fact_tree_split}
Given an $n$-leaf evolutionary tree $T$ and $k$ disjoint sets
$L_1,\cdots,L_k$ of leaf labels of $T$, the subtrees $T||L_1,\cdots,T||L_k$
can be computed in $O(n)$ time.
\end{fact}
\begin{proof}
The ideas are to preprocess $T$ for answering queries of lowest common
ancestors \cite{HT84, SV88} and to reconstruct subtrees from
appropriate tree traversal numberings~\cite{AHU74, CLR91}.
\end{proof}

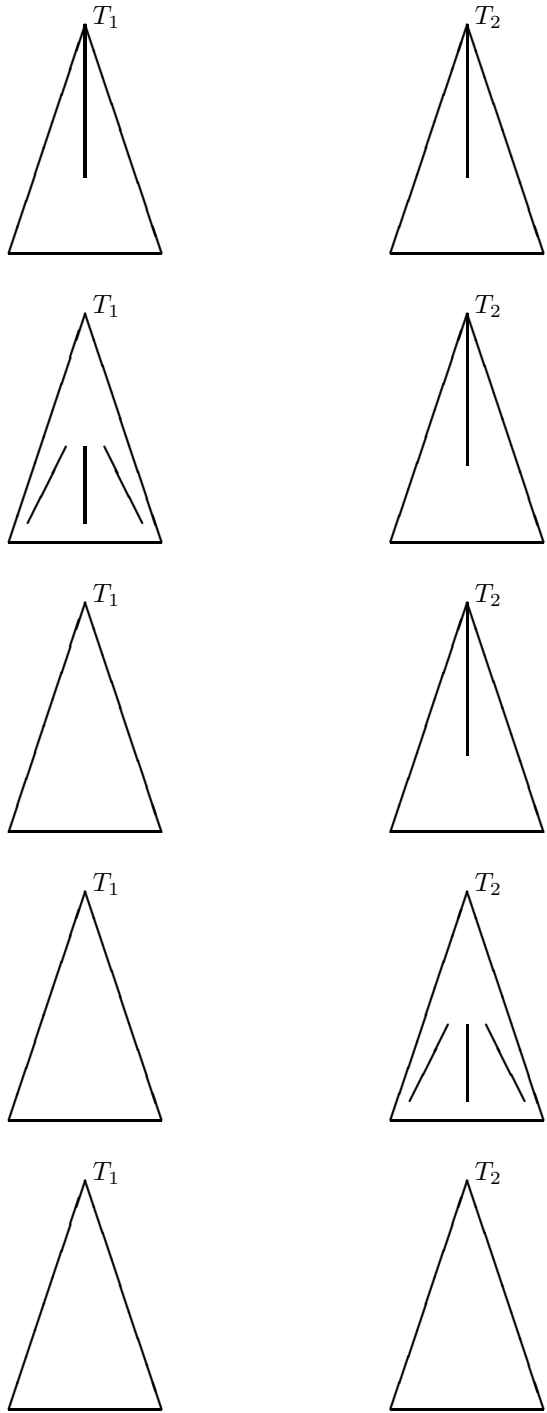
\begin{figure}[p]
\begin{center}

\setlength{\unitlength}{0.4in}
\begin{picture}(2,3.75)(4.0,0)
\thicklines
\put(1.5,0){\line(1,3){1}}
\put(2.5,3){\line(1, -3){1}}
\put(3.5,0){\line(-2, 0){2}}
\put(2.5,3){\line(0, -1){2}}
\put(2.6,3){$T_{1}$}
\put(6.5,0){\line(1,3){1}}
\put(7.5,3){\line(1, -3){1}}
\put(8.5,0){\line(-2, 0){2}}
\put(7.5,3){\line(0,-1 ){2}}
\put(7.6,3){$T_{2}$}
\end{picture}
\end{center}
\begin{center}
\setlength{\unitlength}{0.4in}
\begin{picture}(2,3.75)(4.0,0)
\thicklines
\put(1.5,0){\line(1,3){1}}
\put(2.5,3){\line(1, -3){1}}
\put(3.5,0){\line(-2, 0){2}}
\put(1.75,0.25){\line(1,2){0.5}}
\put(3.25,0.25){\line(-1,2){0.5}}
\put(2.5 ,0.25){\line(0,1){1}}
\put(2.6,3){$T_{1}$}
\put(6.5,0){\line(1,3){1}}
\put(7.5,3){\line(1, -3){1}}
\put(8.5,0){\line(-2, 0){2}}
\put(7.5,3){\line(0,-1 ){2}}
\put(7.6,3){$T_{2}$}
\end{picture}
\end{center}
\begin{center}
\setlength{\unitlength}{0.4in}
\begin{picture}(2,3.75)(4.0,0)
\thicklines
\put(1.5,0){\line(1,3){1}}
\put(2.5,3){\line(1, -3){1}}
\put(3.5,0){\line(-2, 0){2}}
\put(2.6,3){$T_{1}$}
\put(6.5,0){\line(1,3){1}}
\put(7.5,3){\line(1, -3){1}}
\put(8.5,0){\line(-2, 0){2}}
\put(7.5,3){\line(0,-1 ){2}}
\put(7.6,3){$T_{2}$}
\end{picture}
\end{center}
\begin{center}
\setlength{\unitlength}{0.4in}
\begin{picture}(2,3.75)(4.0,0)
\thicklines
\put(1.5,0){\line(1,3){1}}
\put(2.5,3){\line(1, -3){1}}
\put(3.5,0){\line(-2, 0){2}}
\put(2.6,3){$T_{1}$}
\put(6.5,0){\line(1,3){1}}
\put(7.5,3){\line(1, -3){1}}
\put(8.5,0){\line(-2, 0){2}}
\put(6.75,0.25){\line(1,2){0.5}}
\put(8.25,0.25){\line(-1,2){0.5}}
\put(7.5 ,0.25){\line(0,1){1}}
\put(7.6,3){$T_{2}$}
\end{picture}
\end{center}
\begin{center}
\setlength{\unitlength}{0.4in}
\begin{picture}(2,3.75)(4,0)
\thicklines
\put(1.5,0){\line(1,3){1}}
\put(2.5,3){\line(1, -3){1}}
\put(3.5,0){\line(-2, 0){2}}
\put(2.6,3){$T_{1}$}
\put(6.5,0){\line(1,3){1}}
\put(7.5,3){\line(1, -3){1}}
\put(8.5,0){\line(-2, 0){2}}
\put(7.6,3){$T_{2}$}
\end{picture}
\end{center}
\caption{Inputs of Problems~\ref{prob_one_one}--\ref{prob_zero_zero}.}
\label{input_fig}
\end{figure}

Given $T_1$ and $T_2$, our main goal is to evaluate $\rr(T_1,T_2)$
efficiently. Note that $\rr(T_1,T_2)=\rr(T_1||T_2,T_2||T_1)$ and that
$T_1||T_2$ and $T_2||T_1$ can be computed in linear time. Thus, the
remaining discussion assumes that $T_1$ and $T_2$ have the same label set.
To evaluate $\rr(T_1,T_2)$, we actually compute $\ra(T_2,T_1)$ and divide
the discussion among the five problems defined below.  Each problem is
named as a $p$-$q$ case, where $p$ and $q$ are the numbers of tree paths in
$T_1$ and $T_2$ contained in the input.  The inputs of these problems are
illustrated in Figure~\ref{input_fig}.

\begin{problem}[one-one case]\rm
\label{prob_one_one}
\begin{itemize}
\item[] {\INPUT}:
\begin{enumerate}
\item 
$T_1$ and $T_2$;
\item 
root paths $P$ of $T_1$ and $Q$ of $T_2$ with no leaves from their
respective trees;
\item
$\rp(T_1^u,T_2,Q)$ for all $u \in K(P,T_1)$;
\item
$\rp(T_2^v,T_1,P)$ for all $v \in K(Q,T_2)$.
\end{enumerate}
\item[] 
{\OUTPUT}: $\rp(T_1,T_2,Q)$ and $\rp(T_2,T_1,P)$.
\end{itemize}
\end{problem}

The next problem generalizes Problem~\ref{prob_one_one}.

\begin{problem}[many-one case]\rm
\label{prob_many_one}
\begin{itemize}
\item[] {\INPUT}:
\begin{enumerate}
\item 
$T_1$ and $T_2$;
\item 
a tube system $\PP = \{P_1,\cdots,P_m\}$ of $T_1$ and a root path $Q$ of
$T_2$ with no leaf from $T_2$;
\item 
$\rp(T_1^u,T_2,Q)$ for all $P_i$ and $u \in K(P_i,T_1)$;
\item
$\rp(T_2^v,T_1,\PP)$ for all $v \in K(Q,T_2)$.
\end{enumerate}
\item[] {\OUTPUT}: 
\begin{enumerate}
\item
$\rp(T_1^{h_i},T_2,Q)$ for the head $h_i$ of each $P_i$;
\item 
$\rp(T_2,T_1,\PP)$.
\end{enumerate}
\end{itemize}
\end{problem}

\begin{problem}[zero-one case]\rm
\label{prob_zero_one}
\begin{itemize}
\item[] {\INPUT}:
\begin{enumerate}
\item 
$T_1$ and $T_2$;
\item 
a root path $Q$ of $T_2$ with no leaf from $T_2$;
\item
$\ra(T_2^v,T_1)$ for all $v \in K(Q,T_2)$.
\end{enumerate}
\item []
{\OUTPUT}: $\ra(T_2,T_1)$.
\end{itemize}
\end{problem}

The next problem generalizes Problem~\ref{prob_zero_one}.

\begin{problem}[zero-many case]\rm
\label{prob_zero_many}
\begin{itemize}
\item[] {\INPUT}:
\begin{enumerate}
\item 
$T_1$ and $T_2$;
\item 
a tube system $\QQ=\{Q_1,\cdots,Q_m\}$ of $T_2$;
\item 
$\ra(T_2^v,T_1)$ for all $Q_i$ and $v \in K(Q_i,T_2)$.
\end{enumerate}
\item[] {\OUTPUT}: 
$\ra(T_2^{h_i},T_1)$ for the head $h_i$ of each $Q_i$.
\end{itemize}
\end{problem}

Our main goal is to evaluate $\rr(T_1,T_2)$. It suffices to solve the next
problem.

\begin{problem}[zero-zero case]\rm
\label{prob_zero_zero}\rm
\begin{itemize}
\item[] {\INPUT}: $T_1$ and $T_2$.
\item[] {\OUTPUT}: $\ra(T_2,T_1)$.
\end{itemize}
\end{problem}

Our algorithms for these problems are called {\it One-One, Many-One,
  Zero-One, Zero-Many} and {\it Zero-Zero}, respectively.  Each algorithm
except One-One uses the preceding one in this list as a subroutine. These
reductions are based on the rake operation defined in~\S\ref{sec_rake}.  We
give One-One in~\S\ref{sec_one_one} and the other four
in~\S\ref{sec_many_one}-\ref{sec_zero_zero}.

These five algorithms assume that the input trees $T_1$ and $T_2$ have $n$
leaves each and $d$ is the maximum degree.  We use integer sort and radix
sort {\cite{AHU74, CLR91}} extensively to help achieve the desired time
complexity.  (For brevity, from here onwards, radix sort refers to both
integer and radix sorts.)  For this reason, we make the following {\it
integer indexing assumptions}:
\begin{itemize}
\item
An integer array of size $O(n)$ is allocated to each algorithm.
\item
The vertices of $T_1$ and $T_2$ are indexed by integers from $[1,O(n)]$.
\item
The leaf labels are indexed by integers from $[1,O(n)]$.
\end{itemize}

We call Zero-Zero only once to compare two given trees. Consequently, we
may reasonably assume that the tree vertices are indexed with integers from
$[1,O(n)]$.  When we call Zero-Zero, we simply allocate an array of size
$O(n)$. As for indexing the leaf labels, this paper considers only
evolutionary trees whose leaf labels are drawn from a total order.  Before
we call Zero-Zero, we can sort the leaf labels and index them with integers
from $[1,O(n)]$. This preprocessing takes $O(n\log n)$ time, which is well
within our desired time complexity for Zero-Zero.

The other four algorithms are called more than once, and their integer
indexing assumptions are maintained in slightly different situations from
that for Zero-Zero.  When an algorithm issues subroutine calls, it is
responsible for maintaining the indexing assumptions for the callees. In
certain cases, the caller uses radix sort to reindex the labels and the
vertices of each callee's input trees. The caller also partitions its array
into segments and allocates to each callee a segment in proportion to that
callee's input size.  The new indices and the array segments for subroutine
calls can be computed in obvious manners within the desired time complexity
of each caller.  For brevity of presentation, such preprocessing steps are
omitted in the descriptions of the five algorithms.

Some inputs to the algorithms are mappings.  We represent a mapping $f$ by
the set of all pairs $(x,f(x))$. With this representation, the total size
of the input mappings in an algorithm is $O(n)$.  Since the input mappings
have integer values at most $n$, this representation and the integer
indexing assumptions together enable us to evaluate the input mappings at
many points in a batch by means of radix sort.  Other mappings that are
produced within the algorithms are similarly evaluated.  When these
algorithms are detailed, it becomes evident that such evaluations can
computed in straightforward manners in time linear in $n$ and the number of
points evaluated.  The descriptions of these algorithms assume that the
values of mappings are accessed by radix sort.

\section{The rake-based reductions}\label{sec_reductions}
For ease of understanding, our solutions to
Problems~\ref{prob_one_one}--\ref{prob_zero_zero} are presented in a
different order from their logical one.  This section assumes the following
theorem for Problem~\ref{prob_one_one} and uses it to solve
Problems~\ref{prob_many_one}--\ref{prob_zero_zero}.  In
\S\ref{subsec_alg_one_one}, we prove this theorem by giving an algorithm,
called {\it One-One}, that solves Problem~\ref{prob_one_one} within the
theorem's stated time bounds.

\begin{theorem}\label{thm_one_one_primary}
  Problem~\ref{prob_one_one} can be solved in
  \(O(nd^2\log{d}+n\log(p+1)\log(q+1))\) time or alternatively in
  \(O(nd\sqrt{d}\log{n}+n\log(p+1)\log(q+1))\) time.
\end{theorem}
\begin{proof}
Follows from Theorem~\ref{thm_one_one} at the end of
\S\ref{subsec_alg_one_one}.   
\end{proof}

\subsection{The many-one case}\label{sec_many_one}
The following algorithm is for Problem~\ref{prob_many_one} and uses One-One
as a subroutine. Note that Problem~\ref{prob_many_one} is merely a
multi-path version of Problem~\ref{prob_one_one}.

\noindent
{\ALGORITHM} Many-One;
\newline
{\BEGIN}
\begin{enumerate}
\item\label{alg_many_one_split}
For all $P_i$, compute $T_{1,i} = T_1^{h_i}$, $T_{2,i} = T_2||T_{1,i}$, and
$Q_i=Q||T_{1,i}$;
\item\label{alg_many_one_empty}
For all empty $Q_i$, compute part of the output as follows:
\begin{enumerate}
\item\label{alg_many_one_root_1}
Compute the root ${\hat{v}}$ of $T_{2,i}$ and $v \in K(Q,T_2)$ such that
${\hat{v}} \in T_2^{v}$;
\item\label{alg_many_one_eval_1}
$\rp(T_1^{h_i},T_2,Q)(\hat{v})\leftarrow\rp(T_2^v,T_1,\PP)(h_i)$;
(\remark. $H(Q_i,T_{2,i})=\{\hat{v}\}$. This is part of the output.)
\item\label{alg_many_one_eval_2}
For all $x \in H(P_i,T_1)$, $\rp(T_2,T_1,\PP)(x)
\leftarrow \rp(T_2^{v},T_1,\PP)(x)$;
(\remark. This is part of the output.)
\end{enumerate}
\item\label{alg_many_one_nonempty}
For all nonempty $Q_i$, compute the remaining output as follows: (\remark.
The many-one case is reduced to the one-one case with input $T_{1,i}$,
$T_{2,i}$, $P_i$~and~$Q_i$.)
\begin{enumerate}
\item\label{alg_many_one_eval_3}
For all $u \in K(P_i,T_{1,i})$, $\rp(T^u_{1,i},T_{2,i},Q_i) \leftarrow
\rp(T_1^u,T_2,Q)$;
\item\label{alg_many_one_eval_4}
For all ${\hat{v}} \in K(Q_i,T_{2,i})$, compute
$\rp(T_{2,i}^{\hat{v}},T_{1,i},P_i)$ as follows:
\begin{enumerate}
\item\label{alg_many_one_root_2}
Compute the vertex $v \in K(Q,T_2)$ such that ${\hat{v}} \in T_2^{v}$;
\item\label{alg_many_one_eval_5}
$\rp(T^{\hat{v}}_{2,i},T_{1,i},P_i)(x)
\leftarrow
\rp(T_2^{v},T_1,\PP)(x)$ 
for all $x \in H(P_i,T_{1,i})$;
\end{enumerate}
\item\label{alg_many_one_subcall}
Compute $\rp(T_{1,i},T_{2,i},Q_i)$ and $\rp(T_{2,i},T_{1,i},P_i)$
by applying One-One to $T_{1,i}, T_{2,i}$, $P_i$, $Q_i$ and the mappings
computed at Steps~\ref{alg_many_one_eval_3} and
{\ref{alg_many_one_eval_4}};
\item\label{alg_many_one_eval_6}
$\rp(T_1^{h_i},T_2,Q)\leftarrow \rp(T_{1,i},T_{2,i},Q_i)$; (\remark. This is
part of the output.)
\item\label{alg_many_one_eval_7}
For all $x \in H(P_i,T_{1,i})$, $\rp(T_2,T_1,\PP)(x) \leftarrow
\rp(T_{2,i},T_{1,i},P_i)(x)$; (\remark.  This is part of the output.)
\end{enumerate}
\end{enumerate}
{\END}.

\begin{theorem}\label{thm_many_one}
Many-One solves Problem~\ref{prob_many_one} with the following time
complexities:
\[O(nd^2\log{d}+\log(1+t(Q)){\cdot}\sum_{i=1}^{m}b(P_i,T_1)\log(1+t(P_i))),\]
or alternatively
\[O(nd\sqrt{d}\log{n}+
\log(1+t(Q)){\cdot}\sum_{i=1}^{m}b(P_i,T_1)\log(1+t(P_i))).\]
\end{theorem}
\begin{proof}
Since $T_1$ and $T_2$ have the same label set, all $T_{2,i}$ are nonempty.
To compute the output $\rp$, there are two cases depending on whether
$Q_i$ is empty or nonempty. These cases are computed by
Steps~\ref{alg_many_one_empty} and {\ref{alg_many_one_nonempty}}. The
correctness of Many-One is then determined by that of
Steps~\ref{alg_many_one_eval_1},
\ref{alg_many_one_eval_2},
\ref{alg_many_one_eval_3}, \ref{alg_many_one_eval_4},
\ref{alg_many_one_eval_5}, \ref{alg_many_one_eval_6}  and 
\ref{alg_many_one_eval_7}. These steps can be verified using 
Lemma~\ref{lem_tree_restrict}.  As for the time complexity, these steps
take $O(n)$ time using radix sort to evaluate $\rp$.
Step~\ref{alg_many_one_split} uses Fact~\ref{fact_tree_split} and takes
$O(n)$ time.  Steps~\ref{alg_many_one_root_1} and
{\ref{alg_many_one_root_2}} take $O(n)$ time using tree traversal and radix
sort.  As discussed in \S\ref{sec_tree_def},
Step~\ref{alg_many_one_subcall} preprocesses the input of its One-One calls
to maintain their integer indexing assumptions. We reindex the labels and
vertices of $T_{1,i}$ and $T_{2,i}$ and pass the new indices to the calls.
We also partition Many-One's $O(n)$-size array to allocate a segment of
size $|T_{1,i}|$ to the call with input $T_{1,i}$.  Since the total input
size of the calls is $O(n)$, this preprocessing takes $O(n)$ time in an
obvious manner.  After this preprocessing, the running time of
Step~\ref{alg_many_one_subcall} dominates that of Many-One. The stated time
bounds follow from Theorem~\ref{thm_one_one_primary} and the fact that
$Q_i$ is not 
longer than $Q$ and the degrees of $T_{2,i}$ are at most $d$.
\end{proof}

\subsection{The zero-one case}\label{sec_zero_one}
The following algorithm is for Problem~\ref{prob_zero_one}. It uses
Many-One as a subroutine to recursively compare $T_2$ with the subtrees of
$T_1$ rooted at the heads of the tubes obtained by iteratively raking
$T_1$. The tubes obtained by the first rake are compared with $T_2$ first,
and the tube obtained by the last rake is compared last.

\noindent
{\ALGORITHM} Zero-One; \newline {\BEGIN}
\begin{enumerate}
\item\label{alg_many_one_trivial_1}
$S \leftarrow T_1$;
\item\label{alg_many_one_trivial_2}
$LF \leftarrow \ltube(S)$; ({\remark}. $LF$ consists of the leaves of
$T_1$.)
\item\label{alg_zero_one_eval_1}
For all $x \in LF$, $\ra(T_2,T_1)(x) \leftarrow 1$; (\remark. This is part
of the output.)
\item\label{alg_zero_one_init}\label{alg_zero_one_eval_2}
For all $u \in LF$, $\rp(T_1^u,T_2,Q)(y) \leftarrow 1$, where $y$ is
the unique vertex of $T_2||T_1^u$; ({\remark}. This is the base case of
rake-based recursion.)
\item\label{alg_many_one_trivial_3}
$S \leftarrow S - \ltube(S)$;
\item\label{alg_zero_one_loop}
{\WHILE} $S$ is not empty {\DO} the following steps:
\begin{enumerate}
\item\label{alg_zero_one_tube}\label{alg_many_one_trivial_4}
Compute $\ltube(S) = \{P_1,\cdots,P_m\}$;
\item \label{alg_zero_one_map_1}\label{alg_many_one_trivial_5}
Gather the mappings $\rp(T_1^u,T_2,Q)$ for all $P_i$ and $u
\in K(P_i,T_1)$; 
(\remark. These mappings are either initialized at
Step~\ref{alg_zero_one_init} or computed at previous iterations of
Step~\ref{alg_zero_one_iterate}.)
\item\label{alg_zero_one_map_2}\label{alg_zero_one_eval_3}
$\rp(T_2^v,T_1,\ltube(S))(x) \leftarrow \ra(T_2^v,T_1)(x)$ for all $v
\in K(Q,T_2)$ and $x \in \cup_{i=1}^m H(P_i,T_1)$;
\item\label{alg_zero_one_iterate}
Compute $\rp(T_1^{h_i},T_2,Q)$ for the head $h_i$ of each $P_i$ and
$\rp(T_2,T_1,\ltube(S))$ by applying Many-One to $T_1$, $T_2$,
$\ltube(S)$, $Q$ and the mappings obtained at
Steps~\ref{alg_zero_one_map_1} and {\ref{alg_zero_one_map_2}};
(\remark. This is the recursion step of rake-based recursion.)
\item\label{alg_zero_one_eval_4}
For all $x \in \cup_{i=1}^m K(P_i,T_1),
\ra(T_2,T_1)(x) \leftarrow \rp(T_2,T_1,\ltube(S))(x)$;
(\remark. This is part of the output.)
\item\label{alg_many_one_trivial_6}
$S \leftarrow S - \ltube(S)$;
\end{enumerate}
\end{enumerate}
{\END}.

\begin{theorem}\label{thm_zero_one}
Zero-One solves Problem~\ref{prob_zero_one} with the following time
complexities:
\[O(nd^2\log{d}\log{n}+n\log{n}\log(1+t(Q))),\]
or alternatively
\[O(nd\sqrt{d}\log^2{n}+n\log{n}\log(1+t(Q))).\]
\end{theorem}
\begin{proof}
The $\ltube(S)$ at Step~\ref{alg_zero_one_tube} is a tube system. The heads
of the tubes in $\ltube(S)$ become children of the tubes in future
$\ltube(S)$.  The vertices $u \in K(P_i,T_1)$ at
Step~\ref{alg_many_one_trivial_5} are either leaves of $T_1$ or heads of
the tubes in previous $\ltube(S)$.  These properties ensure the correctness
of the rake-based recursion. The remaining correctness proof uses
Lemma~\ref{lem_tree_restrict} to verify the correctness of Steps
{\ref{alg_zero_one_eval_1}}, {\ref{alg_zero_one_eval_2}},
{\ref{alg_zero_one_eval_3}} and {\ref{alg_zero_one_eval_4}}.  Steps
{\ref{alg_many_one_trivial_1}-\ref{alg_many_one_trivial_3}},
\ref{alg_many_one_trivial_4},
\ref{alg_many_one_trivial_5} and
\ref{alg_many_one_trivial_6} are straightforward and take $O(n)$ time.
Step {\ref{alg_zero_one_eval_3}} and {\ref{alg_zero_one_eval_4}} take
$O(n)$ time using radix sort to access $\rp$ and $\ra$.  At
Step~\ref{alg_zero_one_iterate}, to maintain the integer indexing
assumptions for the call to Many-One, we simply pass to Many-One the
indices of $T_1$ and $T_2$ and the whole array of Zero-One.
Step~\ref{alg_zero_one_iterate} has the same time complexity as Zero-One.
The desired time bounds follow from Theorems~\ref{thm_rake} and
Theorem~\ref{thm_many_one}.
\end{proof}

\subsection{The zero-many case}\label{sec_zero_many}
The following algorithm is for Problem~\ref{prob_zero_many} and uses
Zero-One as a subroutine.  Note that Problem~\ref{prob_zero_many} is merely
a multi-path version of Problem~\ref{prob_zero_one}.

\noindent
{\ALGORITHM} Zero-Many;
\newline
{\BEGIN}
\begin{enumerate}
\item
For all $Q_i$, compute $T_{2,i} = T_2^{h_i}$ and $T_{1,i} = T_1||T_{2,i}$;
\item\label{alg_zero_many_map}
For all $Q_i$ and $v \in K(Q_i,T_{2,i})$, $\ra(T_{2,i}^v,T_{1,i})
\leftarrow \ra(T_2^v,T_1)$;
\item 
For all $Q_i$, compute $\ra(T_{2,i},T_{1,i})$ by applying Zero-One to
$T_{1,i}, T_{2,i}$, $Q_i$ and the mapping computed at
Step~\ref{alg_zero_many_map};
\item
For all $Q_i, \ra(T_2^{h_i},T_1) \leftarrow \ra(T_{2,i},T_{1,i})$;
(\remark. This is the output.)
\end{enumerate}
{\END}.

\begin{theorem}\label{thm_zero_many}
Zero-Many solves Problem~\ref{prob_zero_many} with the following time
complexities:
\[O(nd^2\log{d}\log{n}+\log{n}{\cdot}\sum_{i=1}^{m}b(Q_i,T_2)\log(1+t(Q_i))),\]
or alternatively
\[O(nd\sqrt{d}\log^2{n}+\log{n}{\cdot}\sum_{i=1}^{m}b(Q_i,T_2)\log(1+t(Q_i))).\]
\end{theorem}
\begin{proof}
The proof is similar to that of Theorem~\ref{thm_many_one}.  The time
bounds follow from Theorem~\ref{thm_zero_one}.
\end{proof}

\subsection{The zero-zero case}\label{sec_zero_zero}
The following algorithm is for Problem~\ref{prob_zero_zero}. It uses
Zero-Many as a subroutine to recursively compare $T_1$ with the subtrees of
$T_2$ rooted at the heads of the tubes obtained by iteratively raking
$T_2$.  The tubes obtained by the first rake are compared with $T_1$ first,
and the tube obtained by the last rake is compared last.

\noindent
{\ALGORITHM} Zero-Zero;
\newline
{\BEGIN}
\begin{enumerate}
\item
$S \leftarrow T_2$;
\item
$LF \leftarrow \ltube(S)$; (\remark. $LF$ consists of the leaves of $T_2$.)
\item\label{alg_zero_zero_init}
For all $v \in LF$, $\ra(T_2^v,T_1)(x) \leftarrow 1$, where $x$ is the
only vertex in $T_1||T_2^v$; (\remark. This is the base case of rake-based
recursion.)
\item
$S \leftarrow S - \ltube(S)$;
\item\label{alg_zero_zero_loop}
{\WHILE} $S$ is not empty {\DO}
\begin{enumerate}
\item
Compute $\ltube(S) = \{Q_1,\cdots,Q_m\}$;
\item\label{alg_zero_zero_map}
Gather the mappings $\ra(T_2^v,T_1)$ for all $Q_i$ and $v
\in K(Q_i,T_2)$; 
(\remark. These mappings are either initialized at
Step~\ref{alg_zero_zero_init} or computed at previous iterations of
Step~\ref{alg_zero_zero_induction}.)
\item\label{alg_zero_zero_induction}
Compute $\ra(T_2^{h_i},T_1)$ for the head $h_i$ of each $Q_i$ by applying
Zero-Many to $T_1, T_2, \ltube(S)$ and the mappings obtained at
Step~\ref{alg_zero_zero_map}.  (\remark.  This is the recursion step of
rake-based recursion.)
\item
$S \leftarrow S - \ltube(S)$;
\end{enumerate}
\item
$\ra(T_2,T_1) \leftarrow \ra(T_2^h,T_1)$, where $h$ is the root of $T_2$;
(\remark. This is the output.  If $T_2$ has only one vertex,
$\ra(T_2^h,T_1)$ is computed at Step~\ref{alg_zero_zero_init}; otherwise
it is computed at the last iteration of
Step~\ref{alg_zero_zero_induction}.)
\end{enumerate}
{\END}.

\begin{theorem}\label{thm_zero_zero}
Zero-Zero solves Problem~\ref{prob_zero_zero} within
\(O(nd^2\log{d}\log^2{n})\) time or alternatively within 
\(O(nd\sqrt{d}\log^3{n})\) time.
\end{theorem}
\begin{proof}
The proof is similar to that of Theorem~\ref{thm_zero_one}.  The time
bounds follow from Theorems~\ref{thm_rake} and \ref{thm_zero_many}.
\end{proof}

\section{The one-one case}\label{sec_one_one}
Our algorithm for Problem~\ref{prob_one_one} makes extensive use of
bisection-based dynamic programming and implicit computation in compressed
formats.  This problem generalizes the longest common subsequence problem
\cite{AG87, Goldberg82, Hirschberg75, Hirschberg77, HS77}, which has
efficient dynamic programming solutions.  A direct dynamic programming
approach to our problem would recursively solve the problem with $T_1^x$
and $T_2^y$ in place of $T_1$ and $T_2$ for all vertices $x \in P$ and $y
\in Q$.  This approach may require solving $\Omega(n^2)$ subproblems.  To
improve the time complexity, observe that the number of leaves in a maximum
agreement subtree of $T_1^x$ and $T_2^y$ can range only from 0 to $n$.
Moreover, this number never increases when $x$ moves from the root of $T_1$
along $P$ to $P$'s endpoint, and $y$ remains fixed, or vice versa.
Compared to the length of $P$, $\rr(T_1^x,T_2^y)$ often assumes relatively
few different values.  Thus, to compute this number along $P$, it is useful
to compute the locations at $P$ where the number decreases.  We can find
those locations with a bisection scheme and use them to implicitly solve
the $O(n^2)$ subproblems in certain compressed formats.  We first describe
basic techniques used in such implicit computation in \S\ref{sec_condensed}
and then proceed to discuss bisection-based dynamic programming techniques
in \S\ref{sec_simplify}--\S\ref{subsec_recur}. We combine all these
techniques to give an algorithm to solve Problem~\ref{prob_one_one} in
\S\ref{subsec_alg_one_one}.

\subsection{Condensed sequences}\label{sec_condensed}
For integers $k_1$ and $k_2$ with $k_1 \leq k_2$, let
$[k_1,k_2]=\{k_1,\cdots,k_2\}$, i.e., the integer interval between $k_1$
and $k_2$. The {\it length} of an integer interval is the number of its
integers.  The {\it upper} and {\it lower halves} of an even length
$[k_1,k_2]$ are $[\uhalf{k_1}{k_2}]$ and $[\lhalf{k_1}{k_2}]$,
respectively.  The {\it regular} integer intervals are defined recursively.
For all integers $\alpha \geq 0$, $[1,2^\alpha]$ is regular. The upper and
lower halves of an even length regular interval are also regular.  

For example, $[1,8]$ is regular.  Its regular subintervals are $[1,4]$,
$[5,8]$, $[1,2]$, $[3,4]$, $[5,6]$, $[7,8]$, and the singletons $[1,1],
[2,2], \ldots, [8,8]$.

A {\it normal sequence} is a nonincreasing sequence $\{f(j)\}_{j=1}^l$ of
nonnegative numbers.  A normal sequence is {\it nontrivial} if it has at
least one nonzero term. 

For example, $5, 4, 4, 0$ is a nontrivial normal sequence, whereas $0, 0
,0$ is a trivial one.

Let $f_1,\cdots,f_k$ be $k$ normal sequences of length $l$.  An {\it
interval query} for $f_1,\cdots,f_k$ is a pair $([k_1,k_2],j)$ where
$[k_1,k_2] \subseteq [1,k]$ and $j \in [1,l]$.  If $k_1=k_2$,
$([k_1,k_2],j)$ is also called a {\it point query}.  The {\it value} of a
query $([k_1,k_2],j)$ is $\max_{k_1 \leq i \leq k_2}f_i(j)$. A query
$([k_1,k_2],j)$ is {\it regular} if $[k_1,k_2]$ is a regular integer
interval.

For example, let
\[\begin{array}{lcl}
  f_1 & = & 5, 4, 4, 3, 2; \\ 

  f_2 & = & 8, 7, 4, 2, 0; \\ 

  f_3 & = & 9, 9, 5, 0, 0.
\end{array}\]
Then, $f_1$, $f_2$ and $f_3$ are normal sequences of length 5.  Here, $k=3$
and $l=5$. Thus, $([1,3],2)$ is an interval query; its value is
$\max\{f_1(2),f_2(2),f_3(2)\}=9$.  The pair $([1,1],3)$ is a point query;
its value is $f_1(3)=4$. The pair $([1,2],2)$ is a regular query; its
values is $\max\{f_1(2),f_2(2)\} = 7$.

The {\it joint} of $f_1,\cdots,f_k$ is the normal sequence $\hat{f}$ also
of length $l$ such that $\hat{f}(j)=\max\{f_1(j),\cdots,f_k(j)\}$.

Continuing the above example, the joint of $f_1,f_2,f_3$ is
\[\begin{array}{lcl}
  \hat{f} & = & 9,9,5,3,2.
\end{array}\]

The {\it minimal condensed form} of a normal sequence $\{f(j)\}_{j=1}^l$ is
the set of all pairs $(j,f(j))$ where $f(j) \neq 0$ and $j$ is the largest
index of any $f(j')$ with $f(j')=f(j)$.  A {\it condensed form} is a set of
pairs $(j,f(j))$ that includes the minimal condensed form.  The {\it size}
of a condensed form is the number of pairs in it.  The {\it total size} of
a collection of condensed forms is the sum of the sizes of those forms.

Continuing the above example, the minimal condensed form of $f_3$ is
$\{(2,9),(3,5)\}$; its size is 2. The set $\{(1,9),(2,9),(3,5),(5,0)\}$ is
a condensed form of $f_3$; its size is 4.  The total size of these two
forms is 6.

\begin{lemma}\label{lem_joint}
Let $F_1,\cdots,F_k$ be sets of nontrivial normal sequences of length $l$.
Let $\hat{f}_i$ be the joint of the sequences in $F_i$.  Given a condensed
form of each sequence in each $F_i$, we can compute the minimal condensed
forms of all $\hat{f}_i$ in $O(l+s)$ time where $s$ is the total size of
the input forms.
\end{lemma}
\begin{proof} 
The desired minimal forms can be computed by the two steps below:
\begin{enumerate}
\item
Sort the pairs in the given condensed forms for $F_i$ into a sequence in
the increasing order of the first components of these pairs.
\item
Go through this sequence to delete all unnecessary pairs to obtain the
minimal condensed form of $\hat{f}_i$.
\end{enumerate}
We can use radix sort to implement Step 1 in $O(l+s)$ time for all $F_i$.
Step 2 can be easily implemented in $O(s)$ time for all $F_i$.
\end{proof}

\begin{lemma}\label{lem_query}
Let $f_1,\cdots,f_k$ be nontrivial normal sequences of length $l$.  Assume
that the input consists of a condensed form of each $f_i$ with a total size
of $s$.
\begin{enumerate}
\item 
We can evaluate $m$ point queries in $O(m+l+s)$ time.
\item
We can evaluate $m_1$ regular queries and $m_2$ irregular queries in a
total of $O(m_1+(m_2+l+s)\log(k+1))$ time.
\end{enumerate}
\end{lemma}
\begin{proof}
The proof of Statement 1 uses radix sort in an obvious manner.  To prove
Statement 2, we assume without loss of generality that $k$ is a power of
two.  The input queries can be evaluated by the following three stages
within the desired time bound.

Stage 1. For each regular interval $[k_1,k_2] \subseteq [1,k]$, let
$f[k_1,k_2]$ be the joint of $f_{k_1},\cdots,f_{k_2}$.  We use
Lemma~\ref{lem_joint} $O(\log(k+1))$ times to compute the minimal condensed
forms of all $f[k_1,k_2]$.  The total size of these forms is
$O(s\log(k+1))$.  This stage takes $O((l+s)\log(k+1))$ time.

Stage 2. For each irregular input query $([i_1,i_2],j)$, we partition
$[i_1,i_2]$ into $O(\log(k+1))$ regular subintervals
$[h_1,h_2],[h_2+1,h_3], \cdots, [h_{r-1}+1,h_r]$.  Then, the value of
$([i_1,i_2],j)$ is the maximum of those of
$([h_1,h_2],j),\cdots,([h_{r-1}+1,h_r],j)$. These regular queries are point
queries for $f[h_1,h_2],\cdots,f[h_{r-1}+1,h_r]$.  Together with the given
$m_1$ regular queries, we have now generated $O(m_1+m_2\log(k+1))$ point
queries for all $f[k_1.k_2]$.  This stage takes $O(m_1+m_2\log(k+1))$ time.

Stage 3.  We use Statement 1 and the minimal condensed forms of
$f[k_1.k_2]$ to evaluate the points queries generated at Stage 2.  Once the
values of these point queries are obtained, we can easily compute the
values of the input queries. This stage takes
$O(m_1+m_2\log(k+1)+l+s\log(k+1))$ time.
\end{proof} 

\subsection{Normalizing the input}\label{sec_simplify} 
To solve Problem~\ref{prob_one_one}, we first augment its input $T_1, T_2,
P$ and $Q$ in order to simplify our discussion.  Let $P=x_1,\cdots,x_p$ and
$Q=y_1,\cdots,y_q$.  Without loss of generality, we assume that $p \geq q$.
\begin{enumerate}
\item Let $\alpha$ and $\beta$ be the smallest positive integers such that
  $p'=2^\alpha+1$, $q'=2^\beta+1$, $p' \geq q'$, $p' > p$ and $q' > q$.
  (\remark.  The conditions $p'>p$ and $q'>q$ are employed for technical
  simplicity. They can be changed to $p' \geq p$ and $q' \geq q$ with some
  modification on Algorithm One-One.)
\item
Attach to $x_p$ the path $x_{p+1},\cdots,x_{p'}$ and to $y_q$ the path
$y_{q+1},\cdots,y_{q'}$.
\item
Let $P'=x_1,\cdots,x_{p'}$ and $Q'=y_1,\cdots,y_{q'}$.
\item
Attach a leaf to each of $x_{p+1},\cdots,x_{p'-1}$ and
$y_{q+1},\cdots,y_{q'-1}$, two leaves to $x_{p'}$, and two leaves to
$y_{q'}$.
\item
Assign distinct labels to the new leaves which also differ from the
existing labels of $T_1$ and $T_2$.
\item
Let $S_1$ be $T_1$ together with $P'$ and the new leaves of $P'$. Let $S_2$
be $T_2$ together with $Q'$ and the new leaves of $Q'$.
\end{enumerate}

$S_1$ and $S_2$ are evolutionary trees.  $P'$ and $Q'$ contain no leaves
from $S_1$ and $S_2$, and are root paths of these trees.  Let
$n'=\max\{n_1,n_2\}$ where $n_i$ is the number of leaves in $S_i$.  Let
$d'$ be the maximum degree in $S_1$ and $S_2$.

\begin{lemma}\label{lem_aug}
\begin{itemize}
\item
$n'=O(n)$, $p' = O(p)$, $q' = O(q)$, and $d' \leq d+1$.
\item
$\rp(T_1,T_2,Q)=\rp(S_1,S_2,Q')$ and $\rp(T_2,T_1,P)=\rp(S_2,S_1,P')$.
\end{itemize}
\end{lemma}
\begin{proof}
Straightforward.
\end{proof}

In light of Lemma~\ref{lem_aug}, our discussion below mainly works with
$S_1, S_2, P'$ and $Q'$.  Let $G=G_P \cup G_Q$ where $G_P$ is the set of
all pairs $(x_i,y_1)$ and $G_Q$ is the set of all $(x_1,y_j)$. To solve
Problem~\ref{prob_one_one}, a main task is to evaluate $\rr(S_1^x,S_2^y)$
for $(x,y) \in G$.  The output $\rp$ values that are excluded here can be
retrieved directly from the input $\rp$ mappings.

\subsection{Predecessors}\label{subsec_predecessor}
A pair $(x_{i'},y_{j'})$ is a {\it predecessor} of a distinct $(x_i,y_j)$
if $i\leq i'$ and $j \leq j'$.  One-One proceeds by recursively reducing
the problem of computing $\rr(S_1^x,S_2^y)$ to that of computing the $\rr$
values of the {\it $P$-predecessor, $Q$-predecessor and $PQ$-predecessor}
defined below.

Let $P[i,i']$ be the path $x_i,\cdots,x_{i'}$, where $i \leq i'$.  Let
$X_i$ be the set of the children of $x_i$ in $S_1$ that are not in $P'$.
We similarly define $Q[j,j']$ and $Y_j$.  A pair $(x_i,y_j)$ is {\it
intersecting} if $S_1^u$ and $S_2^v$ have at least one common leaf label
for some $u \in X_i$ and $v
\in Y_j$.  $(P[i,i'],Q[y_j,y_{j'}])$ is {\it intersecting} if some $x_{i''}
\in P[i,i']$ and $y_{i''} \in Q[j,j']$ form an intersecting pair.

The {\it lengths} of $P[i,i']$ and $Q[j,j']$ are those of $[i,i']$ and
$[j,j']$, respectively. A path $P[i,i']$ is {\it regular} if $[i,i']$ is a
regular interval. A {\it regular} $Q[j,j']$ is similarly defined.  We now
construct a tree $\Psi$ over pairs of regular paths; this tree is
slightly different from that of \cite{FT94b}.  The root of $\Psi$ is
$(P[1,p'-1],Q[1,q'-1])$.  A pair $(P[i,i'],Q[j,j']) \in \Psi$ is a leaf if
and only if either (1) $i=i'$, $j=j'$ and $(x_i,y_j)$ is intersecting, or
(2) this pair is nonintersecting.  For a nonleaf
$(P[i,i'],Q[j,j'])\in\Psi$, if $j=j'$, then its children are
$(P[\uhalf{i}{i'}],y_j)$ and $(P[\lhalf{i}{i'}],y_j)$.  Otherwise, this
pair has four children $(P[\uhalf{i}{i'}],Q[\uhalf{j}{j'}])$,
$(P[\uhalf{i}{i'}],Q[\lhalf{j}{j'}])$,
$(P[\lhalf{i}{i'}],Q[\uhalf{j}{j'}])$,
$(P[\lhalf{i}{i'}],Q[\lhalf{j}{j'}])$.

The {\it ceiling of $(P[i,i'],Q[j,j'])$} is $(x_i,y_j)$; its {\it floor} is
$(x_{i'+1},y_{j'+1})$ \cite{FT94b}.  Its {\it $P$-diagonal} is
$(x_{i'+1},y_{j})$; its {\it $Q$-diagonal} is $(x_i,y_{j'+1})$.  Let $E$ be
the set of all ceilings, diagonals, floors of the leaves of $\Psi$.  Let
$B=\{(x_i,y_{q'})~|~i \in [1,p']\}\cup\{(x_{p'},y_j)~|~j \in [1,q']\}$. Due
to its recursive nature, One-One evaluates $\rr(S_1^x,S_2^y)$ for all
$(x,y)
\in G \cup E \cup B$.

Given $(x_i,y_j)$, if $(x_{i+1},y_{i+1}) \in G \cup E \cup B$, then this
pair is the {\it $PQ$-predecessor} of $(x_i,y_j)$.  Let $i'$ be the
smallest index that is larger than $i$ such that $(x_{i'},y_j) \in G \cup E
\cup B$.  This $(x_{i'},y_j)$ is the {\it $P$-predecessor of $(x_i,y_j)$}.
Let $j'$ be the smallest index larger than $j$ such that $(x_i,y_{j'}) \in
G \cup E \cup B$.  This $(x_i,y_{j'})$ is the {\it $Q$-predecessor of
$(x_i,y_j)$}.

\begin{lemma}\label{lem_structure}\
\begin{enumerate}
\item
Each intersecting $(x_i,y_j) \in (G \cup E)-B$ has a $P$-predecessor
$(x_{i+1},y_j)$, a $Q$-predecessor $(x_i,y_{j+1})$ and a $PQ$-predecessor
$(x_{i+1},y_{j+1})$.
\item
Each nonintersecting $(x_i,y_j) \in E-B$ has a $P$-predecessor
$(x_{i'},y_j)$ and a $Q$-predecessor $(x_i,y_{j'})$.  Also,
$(P[i,i'-1],Q[j,j'-1])$ is nonintersecting.
\item
Each nonintersecting $(x_i,y_1) \in G_P-B$ has a $P$-predecessor
$(x_{i+1},y_1)$ and a $Q$-predecessor $(x_i,y_j)$. Moreover,
$(x_i,Q[1,j-1])$ is nonintersecting.
\item
Each nonintersecting $(x_1,y_j) \in G_Q-B$ has a $P$-predecessor $(x_i,y_j)$
and a $Q$-predecessor $(x_1,y_{j+1})$. Moreover, $(P[1,i-1],y_j)$ is
nonintersecting.
\end{enumerate}
\end{lemma}
\begin{proof}
Statement 1 follows from the definitions of $\Psi$ and $E$.  The proofs of
Statements 3 and 4 are similar to Case 3 in the proof of Statement 2 below.

As for Statement 2, by the definition of $B$, $x_{i'}$ and $y_{j'}$ exist.
To show $(P[i,i'-1],Q[j,j'-1])$ is nonintersecting, we consider the
following four cases.  The proofs of their symmetric cases are similar to
theirs and are omitted for brevity.

{\case} 1: $(x_i,y_j)$ is the ceiling of a nonintersecting leaf
$(P[i,i_2],Q[j,j_2]) \in \Psi$.  Since $(x_i,y_{j_2+1})$ and
$(x_{i_2+1},y_j)$ are in $E$, $i' \leq i_2+1$ and $j'\leq j_2+1$. Then
because $(P[i,i_2],Q[j,j_2])$ is nonintersecting, so is
$(P[i,i'-1],Q[j,j'-1])$.

{\case} 2: $(x_i,y_j)$ is the $Q$-diagonal of a nonintersecting leaf
$(P[i,i_2],Q[j_1,j-1])$ (or symmetrically, $(x_i,y_j)$ is the $P$-diagonal
of a nonintersecting leaf $(P[i_1,i-1],Q[j,j_2])$).  Since
$(x_{i_2+1},y_j)$ is the floor of $(P[i,i_2],Q[j_1,j-1])$,
$(x_{i_2+1},y_j)\in E$ and thus $i'
\leq i_2+1$.  Let $j''$ be the smallest index such that $j \leq j''$ and
$(P[i,i_2],y_{j''})$ is intersecting. There are two subcases.

{\case} 2a: $j''$ does not exist.  Then, $(P[i,i_2],Q[j,q'])$ is
nonintersecting and therefore $(P[i,i'-1],Q[j,j'-1])$ is nonintersecting.

{\case} 2b.  $j''$ exists. Let $Q[j_3,j_4]$ be a regular path that contains
$y_{j''}$ and is of the same length as $Q[j_1,j-1]$.  Note that $j
\leq j_3$ and $(P[i,i_2],Q[j_3,j_4])\in\Psi$.  There are two subcases.

{\case} 2b(1): $j_3 = j$.  Then $(x_i,y_j)$ is the ceiling of
$(P[i,i_2],Q[j_3,j_4])$.  Since $(x_i,y_j)$ is nonintersecting, it is the
ceiling of a nonintersecting leaf in $\Psi$ which is a descendant of
$(P[i,i_2],Q[j_3,j_4])$. Therefore, Case 2b(1) is reduced to Case~1.

{\case} 2b(2): $j_3 > j$.  By the construction of $\Psi$, $(x_i,y_{j_3})\in
E$ and thus $j' \leq j_3$.  By the choice of $Q[j_3,j_4]$,
$(P[i,i_2],Q[j,j_3-1])$ is nonintersecting and so is
$(P[i,i'-1],Q[j,j'-1])$.

{\case} 3: $(x_i,y_j)$ is the $Q$-diagonal of an intersecting leaf
$(x_i,y_{j-1})$ (or symmetrically, $(x_i,y_j)$ is the $P$-diagonal of an
intersecting leaf $(x_{i-1},y_j)$). Since $(x_{i+1},y_j) \in E$, $i'=i+1$
and $P[i,i'-1]=x_i$.  Let $j''$ be the smallest index such that $j < j''$
and $(x_i,y_{j''})$ is intersecting. There are two subcases.

{\case} 3a: $j''$ does not exist.  Then, $(x_i,Q[j,q'])$ is nonintersecting
and therefore $(P[i,i'-1],Q[j,j'-1])$ is nonintersecting.

{\case} 3b: $j''$ exists. Then, $(x_i,y_{j''}) \in E$ and $j' \leq j''$.
By the choice of $j''$, $(x_i,Q[j,j''-1])$ is nonintersecting.  Thus,
$(P[i,i'-1],Q[j,j'-1])$ is nonintersecting.

{\case} 4: $(x_i,y_j)$ is the floor of a leaf $(P[i_1,i-1],Q[j_1,j-1])$,
which may or may not be intersecting.  Let $(P[i_3,i_4],Q[j_3,j_4])$ be the
lowest ancestor of $(P[i_1,i-1],Q[j_1,j-1])$ in $\Psi$ such that
$(x_i,y_j)$ is not the floor of $(P[i_3,i_4],Q[j_3,j_4])$.  This ancestor
exists because $(x_i,y_j) \not\in B$.  There are two subcases.

{\case} 4a: $j_3=j_4$ and $i_3 < i_4$.  Then, $P[i_1,i-1]$ is a subpath of
$P[\uhalf{i_3}{i_4}]$ and $i=\frac{i_3+i_4+1}{2}$. Also, $j_3=j_1=j-1$.
Thus, $(x_i,y_j)$ is the $Q$-diagonal of $(P[i,i_4],y_{j-1}) \in \Psi$. By
the construction of $\Psi$, $(x_i,y_j)$ is the $Q$-diagonal of a leaf which
is either $(P[i,i_4],y_{j-1})$ itself or its descendant. Depending on
whether this leaf is nonintersecting or intersecting, Case 4a is reduced to
Case 2 or 3.

{\case} 4b: $j_3 < j_4$ and $i_3 < i_4$. There are two subcases.

{\case} 4b(1): $P[i_1,i-1] \subset P[\uhalf{i_3}{i_4}]$ and $Q[j_1,j-1]
\subset Q[\uhalf{j_3}{j_4}]$.  Note that $i=\frac{i_3+i_4+1}{2}$, $j =
\frac{j_3+j_4+1}{2}$, and $(x_i,y_j)$ is the ceiling of
$(P[\lhalf{i_3}{i_4}],Q[\lhalf{j_3}{j_4}]) \in \Psi$.  Since $(x_i,y_j)$ is
nonintersecting, $(x_i,y_j)$ is the ceiling of a nonintersecting leaf in
$\Psi$ which is $(P[\lhalf{i_3}{i_4}],Q[\lhalf{j_3}{j_4}])$ itself or a
descendant. This reduces Case 4b(1) to Case 1.

{\case} 4b(2): $P[i_1,i-1] \subset P[\uhalf{i_3}{i_4}]$ and $Q[j_1,j-1]
\subset Q[\lhalf{j_3}{j_4}]$ (or symmetrically, $P[i_1,i-1] \subset
P[\lhalf{i_3}{i_4}]$ and $Q[j_1,j-1] \subset Q[\uhalf{j_3}{j_4}]$). Note
that $i=\frac{i_3+i_4+1}{2}$, $j = j_4+1$, and $(x_i,y_j)$ is the
$Q$-diagonal of $(P[\lhalf{i_3}{i_4}],Q[\lhalf{j_3}{j_4}]) \in \Psi$.
Then, $(x_i,y_j)$ is the $Q$-diagonal of a leaf which is
$(P[\lhalf{i_3}{i_4}],Q[\lhalf{j_3}{j_4}])$ itself or a
descendant. Depending on whether this leaf is nonintersecting or
intersecting, Case 4b(2) is reduced to Case 2 or 3.
\end{proof}

\subsection{Counting lemmas}\label{subsec_count}
We now give some counting lemmas that are used
in~\S\ref{subsec_alg_one_one} to bound One-One's time complexity.

For all $(P[i_1,i_2],Q[j_1,j_2]) \in \Psi$,
\begin{itemize}
\item
$C(P[i_1,i_2],Q[j_1,j_2])$ denotes the set of all ceilings of the leaves in
$\Psi$ which are either $(P[i_1,i_2],Q[j_1,j_2])$ itself or its
descendants;
\item
$D(P[i_1,i_2],Q[j_1,j_2])$ denotes the set of all $Q$-diagonals of the
leaves in $\Psi$ which are either $(P[i_1,i_2],Q[j_1,j_2])$ itself or its
descendants;
\item
$I(P[i_1,i_2],Q[j_1,j_2]) = \{(x_i,y_j) ~|~ x_i \in P[i_1,i_2],
y_j \in Q[j_1,j_2]\ \mbox{and}\ (x_i,y_j)$ is intersecting$\}$.
\end{itemize}

\begin{lemma}\label{lem_count_1}
\begin{enumerate}
\item\label{lem_count_1_intersect}
$|I(P[1,p'-1],Q[1,q'-1])| \leq n$.
\item\label{lem_count_large_leaf}
$\Psi$ has $O(n\log(q+1))$ leaves of the form $(P[i_1,i_2],Q[j_1,j_2])$
where ${j_1<j_2}$.
\item\label{lem_count_small_leaf}
$\Psi$ has $O(n\log(q+1))$ pairs of the form $(P[i_1,i_2],y_j)$ where
$P[i_1,i_2]$ is of length $\frac{p'-1}{q'-1} $.
\item\label{lem_count_1_E}
$|E|=O(n\log(p+1))$.
\end{enumerate}
\end{lemma}
\begin{proof} 
Statements 1--3 are proved below.  The proof of Statment 4 is similar to
those of Statements 2 and 3.

Statement 1. For all distinct intersecting pairs $(x_i,y_j)$ and
$(x_{i'},y_{j'})$, the leaf labels shared by the subtrees $T^u_1$ where $u
\in X_i$ and the subtrees $T^v_2$ where $v \in Y_i$ are different from
the shared labels for $X_{i'}$ and $Y_{j'}$.  Statement 1 then follows from
the fact that $S_1$ and $S_2$ share $n$ leaf labels.

Statements 2 and 3.  On each level of $\Psi$, for all distinct pairs
$(P[i_1,i_2],Q[j_1,j_2])$ and $(P[i'_1,i'_2],Q[j'_1,j'_2])$,
$I(P[i_1,i_2],Q[j_1,j_2]) \cap
I(P[i'_1,i'_2],Q[j'_1,j'_2])=\emptyset$. Thus, each level has at most
$|I(P[1,p'-1],Q[1,q'-1])|$ nonleaf pairs. Consequently, from the second level
downwards, each level has at most $4\cdot|I(P[1,p'-1],Q[1,q'-1])|$ pairs.
These two statements then follows from Statement 1 and the fact that the
pairs specified in these two statements are within the top $1+\log(q'-1)$
levels of $\Psi$.
\end{proof}

A pair $(x_i,y_j)$ is {\it $P$-regular} if $[i,i'-1]$ is a regular interval
where $(x_{i'},y_j)$ is the $P$-predecessor of $(x_i,y_j)$.  (We do not
need the notion of $Q$-regular because $p' \geq q'$.)

Given a regular $[i_1,i_2]$, a set $\{h_1,\cdots,h_k\}$ {\it regularly
partitions} $[i_1,i_2]$ if $h_1 = i_1$ and the intervals $[h_1,h_2-1],
[h_2,h_3-1], \cdots, [h_{k-1},h_k-1], [h_k,i_2]$ are all regular.

\begin{lemma}\label{lem_regular}
\begin{enumerate}
\item
Assume that $j > 1$ and $P([i_1,i_2],y_j) \in \Psi$. If the $P$-predecessor
$(x_i,y_j)$ of some $(x_{i'},y_j) \in C(P[i_1,i_2],y_j)$ is not in
$\{(x_{i_2+1},y_j)\} \cup C(P[i_1,i_2],y_j)$, then
$P([i_1,i_2],y_{j-1})\in\Psi$ and $(x_i,y_j) \in D(P[i_1,i_2],y_{j-1})$.
\item
Assume that $j < q'$ and $P([i_1,i_2],y_{j-1}) \in \Psi$. If the
$P$-predecessor $(x_i,y_j)$ of some $(x_{i'},y_j)
\in D(P[i_1,i_2],y_{j-1})$ is not in $\{(x_{i_2+1},y_j) \cup
D(P[i_1,i_2],y_{j-1})$, then $P([i_1,i_2],y_j) \in \Psi$ and $(x_i,y_j)
\in C(P[i_1,i_2],y_j)$.  
\item
For every $(P[i_1,i_2],y_j) \in \Psi$, the set $\{i~|~(x_i,y_j)\in
C(P[i_1,i_2],y_j)\}$ regularly partitions $[i_1,i_2]$ and so does the set
$\{i~|~(x_i,y_j)\in D(P[i_1,i_2],y_j)\}$.
\item
For all $(P[i_1,i_2],y_j)\in\Psi$, every pair in $C(P[i_1,i_2],y_j) \cup
D(P[i_1,i_2],y_j)$ is $P$-regular.
\item
At most $O(n\log(q+1))$ of the nonintersecting pairs of $E$ are
$P$-irregular.
\end{enumerate} 
\end{lemma}
\begin{proof}  
The proofs of Statements 1 and 5 are detailed below.  The proof of
Statement 2 is similar to that of Statement 1 and is omitted.  Statement 3
is obvious.  Statement 4 follows from the first three statements and the
fact that if two sets regularly partition $[i_1,i_2]$, then so does their
union.

Statement 1. Note that $i_1 < i \leq i_2$ and $q' > j > 1$. The pair
$(x_i,y_j)$ can be the ceiling, the $P$-diagonal, the $Q$-diagonal, or the
floor of some leaf $(P[i_3,i_4],Q[j_3,j_4]) \in
\Psi$. These four cases are discussed below.

{\case} 1: $(x_i,y_j)$ is the ceiling.  Then $i=i_3$ and $j=j_3$.  Since
$i_1 < i \leq i_2$ and both $[i,i_4]$ and $[i_1,i_2]$ are regular, $[i,i_4]
\subset [i_1,i_2]$.  Since the length of $P[i_1,i_2]$ is at most
$\frac{p'-1}{q'-1} $, so is the length of $P[i,i_4]$.  Thus
$Q[j_3,j_4]=y_j$ and $(P[i,i_4],y_j)$ is a descendant of
$(P[i_1,i_2],y_j)$. This contradicts the assumption that $(x_i,y_j)\not\in
C(P[i_1,i_2],y_j)$ and this case cannot exist.

{\case} 2: $(x_i,y_j)$ is the $P$-diagonal.  Then $i=i_4+1$ and $j=j_3$.
As in Case~1, $Q[j_3,j_4]=y_j$ and $(P[i_3,i-1],y_j)$ is a descendant of
$(P[i_1,i_2],y_j)$. Thus, there exists a leaf $(P[i,i_6],y_j)$ that is a
descendant of $(P[i_1,i_2],y_j)$.  Because $(x_i,y_j)$ is the ceiling of
this leaf, the existence of this leaf contradicts the assumption that
$(x_i,y_j)\not\in C(P[i_1,i_2],y_j)$ and this case cannot exist.

{\case} 3: $(x_i,y_j)$ is the $Q$-diagonal.  Then, $i=i_3$ and $j=j_4+1$.
As in Case~1, $[i,i_4] \subset [i_1,i_2]$ and $Q[j_3,j_4]=y_{j-1}$. Since
$(P[i,i_4],y_{j-1}) \in \Psi$, $(P[i_1,i_2],y_{j-1}) \in \Psi$. Then
$(P[i,i_4],y_{j-1})$ is a descendant of $(P[i_1,i_2],y_{j-1})$ and
$(x_i,y_j) \in D(P[i_1,i_2],y_{j-1})$.

{\case} 4: $(x_i,y_j)$ is the floor.  Then, $i=i_4+1$ and $j=j_4+1$.  As in
Case~3, $(P[i_1,i_2],y_{j-1})\in\Psi$, $Q[j_3,j-1]=y_{j-1}$ and
$(P[i_3,i-1],y_{j-1})$ is a descendant of $(P[i_1,i_2],y_{j-1})$.  Thus,
there is a leaf $(P[i,i_6],y_{j-1})$ which is a descendant of
$(P[i_1,i_2],y_{j-1})$. Since $(x_i,y_j)$ is this leaf's $Q$-diagonal, it
is in $D(P[i_1,i_2],y_{j-1})$.  

Statement 5.  Note that $E$ consists of the following three types of pairs:
\begin{enumerate}
\item
the ceiling, diagonals and floor of a leaf $(P[i_1,i_2],Q[j_1,j_2])\in
\Psi$ where $j_1 < j_2$.
\item
the $P$-diagonal and floor of $(P[i_1,i_2],y_j])\in\Psi$ where $P[i_1,i_2]$
is of length $\frac{p'-1}{q'-1}$.
\item
the pairs in $C(P[i_1,i_2],j]) \cup D(P[i_1,i_2],y_j])$ where
$(P[i_1,i_2],j])\in\Psi$ and $P[i_1,i_2]$ is of length $\frac{p'-1}{q'-1}$.
\end{enumerate}
By Statement 4, only the pairs of the first two types may be
$P$-irregular. This statement then follows from Lemmas
{\ref{lem_count_1}(\ref{lem_count_large_leaf})} and
{\ref{lem_count_1}(\ref{lem_count_small_leaf})}.
\end{proof}

\subsection{Recurrences}\label{subsec_recur}
One-One uses the following formulas to recursively compute
$\rr(S_1^{x_i},S_2^{y_j})$ for $(x_i,y_j) \in G \cup E \cup B$ in terms of
the $\rr$ values of the appropriate $P$-predecessor, $Q$-predecessor and
$PQ$-predecessor of $(x_i,y_j)$.

For vertex subsets $U$ of $S_1$ and $V$ of $S_2$, $\mm(U,V)$ denotes the
maximum weight of any matching of the bipartite graph $(U,V,U{\times}V)$
where the weight of an edge $(u,v)$ is $\rr(S_1^u,S_2^v)$.  Let
$\mm(U,v)=\mm(U,\{v\})$ and $\mm(u,V)=\mm(\{u\},V)$.  Given two vertices $x
\in S_1$ and $y \in S_2$, let $\mmb(U,V,x,y)$ be the maximum weight of any
matching of the same graph without the edge $(x,y)$.

\begin{lemma}\label{lem_recur_B}
For each $(x_i,y_j) \in B$, $\rr(S_1^x,S_2^y)=0$.
\end{lemma}
\begin{proof}
This lemma follows from the fact that $p' > p$, $q > q$ and the new labels
of $S_1$ and $S_2$ are different from one another and the labels of $T_1$
and $T_2$.
\end{proof}

\begin{fact}[\cite{SW93}]\label{fact_recur}
For all vertices $u \in S_1$ and $v \in S_2$,
\[\rr(S_1^u,S_2^v)=
\max\left\{
\begin{array}{l}
\mm(K(u,S_1),K(v,S_2)),\\
\mm(u,K(v,S_2)),\\
\mm(K(u,S_1),v)
\end{array}
\right\}.
\]
\end{fact}
\begin{proof}
To form maximum agreement subtrees of $S_1^u$ and $S_2^v$, there are three
cases.  (1) 
$\mm(K(u,S_1),K(v,S_2))$ accounts for matching $u$ to $v$.
(2) $\mm(u,K(v,S_2))$ accounts for matching $u$ to a proper descendant of
$v$.
(3) $\mm(K(u,S_1),v)$ accounts for matching $v$ to a proper
descendant of $u$.
\end{proof}

\begin{lemma}\label{lem_recur_intersect}
For all $(x_i,y_j)$ where $i < p'$ and $j < q'$, regardless of whether
$(x_i,y_j)$ is intersecting or nonintersecting,
\[
\rr(S_1^{x_i},S_2^{y_j})=\max\left\{
\begin{array}{l}
\mm(X_i,Y_j)+\rr(S_1^{x_{i+1}},S_2^{y_{j+1}}),\\
\mmb(X_i\cup\{x_{i+1}\},Y_j\cup\{y_{j+1}\},x_{i+1},y_{j+1}),\\
\rr(S_1^{x_i},S_2^{y_{j+1}}),\\
\mm(x_i,Y_j),\\
\rr(S_1^{x_{i+1}},S_2^{y_j}),\\
\mm(X_i,y_j)
\end{array}
\right\}.
\]
\end{lemma}
\begin{proof}
This lemma follows from Fact~\ref{fact_recur} with a finer case analysis
for the cases in the proof of Fact~\ref{fact_recur}.
\end{proof}

\begin{lemma}\label{lem_recur_E}
For each nonintersecting $(x_i,y_j) \in E-B$ with $P$-predecessor
$(x_{i'},y_j)$ and $Q$-predecessor $(x_i,y_{j'})$,
\[
\rr(S_1^{x_i},S_2^{y_j})=\max\left\{
\begin{array}{l}
\max_{j'' \in [j,j'-1]} \mm(x_{i'},Y_{j''})
+
\max_{i'' \in [i,i'-1]} \mm(X_{i''},y_{j'}),\\
\rr(S_1^{x_i},S_2^{y_{j'}}), \\
\rr(S_1^{x_{i'}},S_2^{y_j})
\end{array}
\right\}.
\]
\end{lemma}
\begin{proof}
This lemma follows from Lemma~\ref{lem_structure}(2) and is
obtained by iterative applications of Lemma~\ref{lem_recur_intersect}. The
following properties are used.  Since $(P[i,i'-1],Q[j,j'-1])$ is
nonintersecting, for $i'' \in [i,i'-1]$ and $j'' \in [j,j'-1]$,
\begin{itemize}
\item
$\mm(X_{i''},Y_{j''})= 0;$
\item
$\mmb(X_{i''}\cup\{x_{i''+1}\},Y_{j''}\cup\{y_{j''+1}\},x_{i''+1},y_{j''+1})
=\mm(x_{i''},Y_{j''})+\mm(X_{i''},y_{j''});$
\item
$\mm(x_{i''},Y_{j''})=\mm(x_{i'},Y_{j''});$
\item
$\mm(X_{i''},y_{j''})=\mm(X_{i''},y_{j'}).$
\end{itemize}
\end{proof}

For brevity, the symmetric statement of the next lemma for $G_Q$ is
omitted.
\begin{lemma}\label{lem_recur_G}
For all nonintersecting pairs $(x_i,y_1) \in G_P-B$ with
$Q$-predecessor $(x_i,y_j)$,
\[
\rr(S_1^{x_i},S_2^{y_1})=\max\left\{
\begin{array}{l}
\rr(S_1^{x_i},S_2^{y_j}), \\
\rr(S_1^{x_{i+1}},S_2^{y_1}), \\
\mm(X_i,y_j)+\max_{j' \in [1,j-1]} \mm(x_{i+1},Y_{j'})
\end{array}
\right\}.
\]
\end{lemma}
\begin{proof}
The proof is similar to that of Lemma~\ref{lem_recur_E} and follows from
Lemma~\ref{lem_structure}(3).
\end{proof}

\subsection{The algorithm for Problem~\protect{\ref{prob_one_one}}}
\label{subsec_alg_one_one} 
We combine the discussion
in~\S\ref{subsec_predecessor}--\S\ref{subsec_recur} to give the following
algorithm to solve Problem~\ref{prob_one_one}.

\noindent
{\ALGORITHM} One-One;
\newline
{\BEGIN}
\begin{enumerate}
\item\label{alg_one_one_aug}
Compute $S_1$, $S_2$, $P'$, $Q'$, $\rp(S_1^u,S_2,Q')$ for $u \in
K(P',S_1)$, and $\rp(S_2^v,S_1,P')$ $v \in K(Q',S_2)$;
\item\label{alg_one_one_set}
Compute 
$G \cup E \cup B$,
$B$, 
$I(P[1,p'-1],Q[1,q'-1])-B$,
the set of all nonintersecting pairs in $E-B$, and
the sets of nonintersecting pairs in $G_P-B$ and $G_Q-B$, respectively;
\item\label{alg_one_one_predecessor}
Compute the following predecessors:
\begin{itemize}
\item 
the $P$-predecessor, $Q$-predecessor and $PQ$-predecessor of each pair in
$I(P[1,p'-1],Q[1,q'-1])-B$;
\item
the $P$-predecessor and $Q$-predecessor of each nonintersecting pair in
$E-B$;
\item
the $Q$-predecessor of each nonintersecting pair in $G_P-B$ and the
$P$-predecessor of each nonintersecting pair in $G_Q-B$;
\end{itemize}
\item\label{alg_one_one_nonrr}
For all pairs in $G \cup E \cup B$, compute the non-$\rr$ terms in the
appropriate recurrence formulas of~\S\ref{subsec_recur}:
\begin{itemize}
\item
Lemma~\ref{lem_recur_B} for $B$;
\item
Lemma~\ref{lem_recur_intersect} for $(I(P[1,p'-1],Q[1,q'-1])-B$;
\item
Lemma~\ref{lem_recur_E} for the nonintersecting pairs in $E-B$;
\item
Lemma~\ref{lem_recur_G} for the nonintersecting pairs in $G_P-B$ and its
symmetric statement for the nonintersecting pairs in $G_Q-B$;
\end{itemize}
\item\label{alg_one_one_rr} 
Compute the $\rr(S_1^{x_i},S_2^{y_j})$ for all $(x_i,y_j) \in G \cup E \cup
B$ using the appropriate recurrence formulas given in~\S\ref{subsec_recur}
and the non-$\rr$ terms computed at Step~\ref{alg_one_one_nonrr};
\item\label{alg_one_one_output} 
Compute the output as follows:
\begin{itemize}
\item\label{alg_one_pq_1}
For all $y_j \in Q$, $\rp(T_1,T_2,Q)(y_j) \leftarrow
\rr(S_1^{x_1},S_2^{y_j})$;
\item\label{alg_one_pq_2}
For all $x_i \in P$, $\rp(T_2,T_1,P)(x_i) \leftarrow
\rr(S_1^{x_i},S_2^{y_1})$;
\item\label{alg_one_pqk_1}
For every $v \in K(Q,T_2)$,
$\rp(T_1,T_2,Q)(v)\leftarrow\rp(T_2^v,T_1,P)(h)$ where $h$ is the root of
$T_1||T_2^v$;
\item\label{alg_one_pqk_2}
For every $u\in K(P,T_1)$,
$\rp(T_2,T_1,P)(u)\leftarrow\rp(T_1^u,T_2,Q)(h)$ where $h$ is the root of
$T_2||T_1^u$;
\end{itemize}
\end{enumerate}
{\END}.

To analyze One-One, we first focus on Step~\ref{alg_one_one_nonrr}.  The
recurrences of~\S\ref{subsec_recur} contain only four types of non-$\rr$
terms other than the constant $0$ in Lemma~\ref{lem_recur_B}:
\begin{enumerate}
\item
$\mm(X_i,y_j)$ and $\mm(x_i,Y_j)$;
\item
$\max_{i\in[i_1,i_2]}\mm(X_i,y_j)$ and $\max_{j\in[j_1,j_2]}\mm(x_i,Y_j)$;
\item
$\mm(X_i,Y_j)$;
\item 
$\mmb(X_i\cup\{x_{i+1}\},Y_j\cup\{y_{j+1}\},x_{i+1},y_{j+1})$.
\end{enumerate}

It is important to notice that these non-$\rr$ terms can be simultaneously
evaluated. In light of this observation, we compute these terms by using
the techniques of~\S\ref{sec_condensed} to process the normal sequences
$A_i,A_u,B_j,B_v$ defined below:
\begin{itemize} 
\item
$A_i(j)=\mm(X_i,y_j)$ for all $x_i$ and $y_j$.
\item
$B_j(i)=\mm(x_i,Y_j)$ for all $y_j$ and $x_i$.
\item
$A_u(j)=\rr(S_1^u,S_2^{y_j})$ for all $u \in K(P',S_1)$ and $y_j$.
\item
$B_v(i)=\rr(S_2^v,S_1^{x_i})$ for all $v \in K(Q',S_2)$ and $x_i$.
\end{itemize}

Note that $A_i$ and $A_u$ have length $q'$, and $A_i$ is the joint of all
$A_u$ where $u \in X_i$.  Similarly, $B_j$ and $B_v$ have length $p'$, and
$B_j$ is the joint of all $B_v$ where $v \in Y_j$.

\begin{lemma}\label{lem_mcf}\
\begin{enumerate}
\item
The minimal condensed forms of the sequences $A_u$ and $B_v$ have a
total size of $O(n)$ and can be computed in $O(n)$ time.
\item
The minimal condensed forms of the sequences $A_i$ and $B_j$ have a
total size of $O(n)$ and can be computed in $O(n)$ time.
\end{enumerate}
\end{lemma}
\begin{proof}
Statement 2 follows from Statement 1 and Lemma~\ref{lem_joint}.  Below we
only prove Statement 1 for $A_u$; Statement 1 for $B_v$ is similarly
proved. We first compute a condensed form $\Ab_u$ for each $A_u$ as
follows:
\begin{enumerate}
\item\label{step_mcf_split}
For all $u \in K(P',S_1)$, compute $S_{2,u}=S_2||S_1^u$ and
$Q_u=Q'||S_1^u$.
\item\label{alg_mcf_nonempty}
For all $u$ where $Q_u$ is nonempty, do the following steps:
\begin{enumerate}
\item 
$\Ab_u \leftarrow \{(j,w)~|~y_j \in Q_u, w=\rp(S_1^u,S_2,Q')(y_j)\}$.
\item\label{alg_mcf_bottom}
Compute all tuples $(\hat{v},v,y_j)$ where $\hat{v} \in
K(Q_u,S_{2,u})$, $v \in K(Q',S_2)$, $\hat{v} \in S_2^v$, and $v \in
Y_j$.
\item
Find the smallest $s$ such that some $(\hat{v},v,y_s)$ is obtained at
Step~\ref{alg_mcf_bottom}.
\item
If there is only one $(\hat{v},v,y_s)$, then add to $\Ab_u$ the pair
$(s,w)$ where $w=\rp(S_1^u,S_2,Q')(\hat{v})$.
\end{enumerate}
\item\label{alg_mcf_empty_path}
For all $u$ where $S_{2,u}$ is nonempty and $Q_u$ is empty, do the
following steps:
\begin{enumerate}
\item 
Compute $\hat{v}$, $v$ and $y_s$ where $\hat{v}$ is the root of
$S_{2,u}$, $v \in K(Q',S_2)$, $\hat{v} \in S_2^v$ and $v \in Y_s$.
\item
$\Ab_u\leftarrow\{(s,w)\}$, where $w=\rp(S_1^u,S_2,Q')(\hat{v})$.
\end{enumerate}
\item\label{alg_mcf_empty_tree} 
For all $u$ where $S_{2,u}$ is empty, $\Ab_u\leftarrow\emptyset$.
\end{enumerate}

The correctness proof of this algorithm has three cases.

{\case} 1: $Q_u$ is nonempty.  Let
$y_{j_1},y_{j_2},\cdots,y_{j_k}=Q_u$.  Let $j_0=0$. Then, for all $k'
\in [1,k]$ and all $j \in [j_{k'-1}+1,j_{k'}]$,
$S_2^{y_j}||S_1^u=S_{2,u}^{y_{k'}}$ and by
Lemma~\ref{lem_tree_restrict}, $A_u(j)=\rp(S_1^u,S_2,Q')(y_{k'})$.
There are two subcases for $j > j_k$.

{\case} 1a: Step~\ref{alg_mcf_bottom} finds two or more $(\hat{v},v,y_s)$.
Then $y_s \in Q_u$, $s=j_k$, and for all $j\in[j_k+1,q']$,
$S_2^{y_j}||S_2^u$ is empty and $A_u(j)=0$.

{\case} 1b: Step~\ref{alg_mcf_bottom} finds only one
$(\hat{v},v,y_s)$.  Then $y_s \not\in Q_u$ and $s>j_k$.  For all $j
\in [j_k+1,s], S_2^{y_j}||S_1^u=S_{2,u}^{\hat{v}}$ and 
$A_u(j)=\rp(S_1^u,S_2,Q')(\hat{v})$.  For all
$j\in[s+1,q']$, $S_2^{y_j}||S_1^u$ is empty and $A_u(j)=0$.

Thus, the $\Ab_u$ of Step~\ref{alg_mcf_nonempty} is a condensed form of
$A_u$ for Case 1.

{\case} 2: $S_{2,u}$ is nonempty and $Q_u$ is empty.  This case is similar
to Case 1b, and Step~\ref{alg_mcf_empty_path} computes a correct condensed
form $\Ab_u$ for this case.

{\case} 3: $S_{2,u}$ is empty.  This case is obvious, and
Step~\ref{alg_mcf_empty_tree} correctly computes a condensed form $\Ab_u$
of $A_u$ for this case.

The total size of all $\Ab_u$ is at most that of the $\rp$ mappings of
$S_1, S_2, P'$ and $Q'$, which is the desired $O(n)$.
Step~\ref{step_mcf_split} takes $O(n)$ time using
Fact~\ref{step_mcf_split}.  The other steps can be implemented in $O(n)$
time in straightforward manners using radix sort and tree traversal.  As
discussed in~\S\ref{sec_tree_def}, the $\rp$ mappings are evaluated by
radix sort. Once the forms $\Ab_u$ are obtained, we can in $O(n)$ time
radix sort the pairs in all $\Ab_u$ and then delete all unnecessary pairs
to obtain the desired minimal condensed forms.
\end{proof}

\begin{lemma}\label{lem_nonrr_12}
All the non-$\rr$ terms of the first two types for the pairs in $G
\cup E \cup B$ can be evaluated in $O(n\log(p+1)\log(q+1))$ time.
\end{lemma}
\begin{proof}
The value of $\mm(X_i,y_j)$ is that of the point query $([i,i],j)$ for
$A_1,\cdots,A_{q'}$, and the value of $\max_{i\in[i_1,i_2]}\mm(X_i,y_j)$ is
that of the interval query $([i_1,i_2],j)$.  By
Lemma~\ref{lem_count_1}(\ref{lem_count_1_E}), there are $O(n\log(p+1))$
such terms required for the pairs in $G \cup E \cup B$.  Given the results
of Steps~\ref{alg_one_one_set} and {\ref{alg_one_one_predecessor}} of
One-One, we can determine all such terms and the corresponding queries in
$O(n\log(p+1))$ time.  By Lemma~\ref{lem_regular}(5), only $O(n\log(q+1))$
of these queries are not $P$-regular.  By Lemmas~\ref{lem_mcf}(2) and
{\ref{lem_query}(2)}, we can evaluate these queries in
$O(n\log(p+1)\log(q+1))$ time.  The terms $\mm(x_i,Y_j)$ and
$\max_{j\in[j_1,j_2]}\mm(x_i,Y_j)$ are similarly evaluated is
$O(n\log(p+1)\log(q+1))$ time. The analysis for these terms is easier
because $p' \geq q'$ and it does not involve the notion of $Q$-regularity.
\end{proof}

\begin{lemma}\label{lem_nonrr_34}
The non-$\rr$ terms of the third and the fourth type for the pairs in
$G \cup E \cup B$ can be evaluated within the following time
complexity:
\begin{enumerate}
\item
$O(n d\log d)$ or alternatively $O(n \sqrt{d} \log n)$ for the third
type;
\item
$O(n d^2 \log d)$ or alternatively $O(n d \sqrt{d} \log n)$ for the
fourth type.
\end{enumerate}
\end{lemma}
\begin{proof}
To prove Statement 1, we consider the graphs
$(X_i,Y_j,X_i{\times}Y_j)$ on which the desired terms $\mm(X_i,Y_j)$
are defined.  Let $Z_{i,j}$ be the subgraph of
$(X_i,Y_j,X_i{\times}Y_j)$ constructed by removing all zero-weight
edges and all resulting isolated vertices.  The edges of $Z_{i,j}$ are
computed as follows:
\begin{enumerate}
\item
For all $u \in K(P',S_1)$, compute $S_{2,u}=S_2||S_1^u$ and
$Q_u=Q'||S_1^u$.
\item\label{alg_nonrr_3rd}
For all $S_{2,u}$ is nonempty, do the following steps:
\begin{enumerate}
\item
If $Q_u$ is nonempty, compute all tuples $(u,v,w)$ where $\hat{v}\in
K(Q_u,S_{2,u})$, $v \in K(Q',S_2)$, $\hat{v} \in S_2^v$ and
$w=\rp(S_1^u,S_2,Q')(\hat{v})$.
\item
If $Q_u$ is empty, compute the tuple $(u,v,w)$ where $\hat{v}$ is the root
of $S_{2,u}$, $v \in K(Q',S_2)$, $\hat{v} \in S_2^v$ and
$w=\rp(S_1^u,S_2,Q')(\hat{v})$.
\end{enumerate}
\end{enumerate}
This algorithm captures all the nonzero-weight $(u,v)$.  At
Step~\ref{alg_nonrr_3rd}, $S_{2,u}^{\hat{v}}=S_2^v||S_1^u$ and by
Lemma~\ref{lem_tree_restrict}
$\rr(S_1^u,S_2^v)=\rp(S_1^u,S_2,Q')(\hat{v})$.  Thus, the first two
components of the obtained tuples form the edges of all desired $Z_{i,j}$
and the third components are the weights of these edges.  We use
Fact~\ref{fact_tree_split} to implement Step 1 in $O(n)$ time.  We can
implement Step~\ref{alg_nonrr_3rd} in $O(n)$ time using radix sort and tree
traversal.  Note that Step~\ref{alg_nonrr_3rd} uses radix sort to evaluate
$\rp$ mappings.  With the tuples $(u,v,w)$ obtained, we use radix sort to
construct all desired $Z_{i,j}$ in $O(n)$ time. Let $m_{i,j}$ and $n_{i,j}$
be the numbers of edges and vertices in $Z_{i,j}$, respectively.  Since an
edge weighs at most $n$, we can compute $\mm(X_i,Y_j)$ in
$O(n_{i,j}{\cdot}m_{i,j} + n^2_{i,j}{\cdot}\log n_{i,j})$ and alternatively
in $O(m_{i,j}{\cdot}\sqrt{n_{i,j}}{\cdot}\log(n{\cdot}n_{i,j}))$ time
\cite{GT89, OA92}.  
Statement 1 then follows from the fact that $n_{i,j}\leq2d'$,
$n_{i,j}\leq2m_{i,j}$, and by
Lemma~\ref{lem_count_1}(\ref{lem_count_1_intersect}) the sum of all
$m_{i,j}$ is at most $n$.

To prove Statement 2, we similarly process the bipartite graphs on which the
desired terms $\mmb(X_i\cup\{x_{i+1}\},Y_j\cup\{y_{j+1}\},x_{i+1},y_{j+1})$
are defined.  The key difference from the third type is that in addition to
some of the edges in $Z_{i,j}$, we need certain nonzero-weight
$(u,y_{j+1})$ for $u\in X_i$ and $(x_{i+1},v)$ for $v \in Y_j$.  Since
these edges are required only for intersecting $(x_i,y_j)$, by
Lemma~\ref{lem_count_1}(\ref{lem_count_1_intersect}), $O(dn)$ such edges
are needed.  We use Lemma~\ref{lem_mcf}(1) to compute the weights of these
edges in $O(d n)$ time. Due to these edges, the total time complexity for
the fourth type is $O(d)$ times that for the third type.
\end{proof}

The next theorem serves to prove Theorem~\ref{thm_one_one_primary} given at
the start of \S\ref{sec_reductions}.

\begin{theorem}\label{thm_one_one}
One-One solves Problem~\ref{prob_one_one} with the following time
complexities: 
\[O(nd^2\log{d}+n\log(p+1)\log(q+1)),\] 
or alternatively 
\[O(nd\sqrt{d}\log{n}+n\log(p+1)\log(q+1)).\]
\end{theorem}
\begin{proof}
The correctness of One-One follows from Lemma~\ref{lem_aug} and
{\S\ref{subsec_predecessor}--\S\ref{subsec_recur}}.  As for the time
complexity, Step~\ref{alg_one_one_aug} is obvious and takes $O(n)$ time.
By computing $\Psi$, we can compute the sets $E$ and
$I(P[1,p'-1],Q[1,q'-1])$. Since the leaf labels of $S_1$ and $S_2$ are from
$[1,O(n)]$, each level of $\Psi$ can be computed in $O(n)$ time.  Since
$\Psi$ has $O(\log(p+1))$ levels, $E$ and $I(P[1,p'-1],Q[1,q'-1])$ can be
computed in $O(n\log(p+1))$ time.  With these two sets obtained, we can
compute all the desired sets in $O(n\log(p+1))$ time. Thus,
Step~\ref{alg_one_one_set} takes $O(n\log(p+1))$ time.
Step~\ref{alg_one_one_predecessor} takes $O(n\log(p+1))$ time using radix
sort.  The time complexity of Step~\ref{alg_one_one_nonrr} dominates that
of One-One. This step uses Lemmas~\ref{lem_nonrr_12} and
{\ref{lem_nonrr_34}} and takes $O(n\log(p+1)\log(q+1)+nd^2\log{d})$ time or
alternatively $O(n\log(p+1)\log(q+1)+nd\sqrt{d}\log{n})$ time.
Step~\ref{alg_one_one_rr} spends $O(n\log(p+1))$ time using radix sort to
create pointers from the pairs in $G \cup E \cup B$ to appropriate
predecessors.  Step~\ref{alg_one_one_rr} then takes $O(1)$ time per pair in
$G \cup E \cup B$ and $O(n\log(p+1))$ time in total.
Step~\ref{alg_one_one_output} takes $O(n\log(p+1))$ time. It uses radix
sort to access the desired $\rr$ values and evaluate the input mappings. It
also uses Fact~\ref{fact_tree_split} to compute all $T_1||T_2^v$ and
$T_2||T_1^u$.
\end{proof}

\section{Discussions}\label{sec_disc}
We answer the main problem of this paper with the following theorem and
conclude with an open problem.

\begin{theorem}\label{thm_tree}
  Let $T_1$ and $T_2$ be two evolutionary trees with $n$ leaves each.  Let
  $d$ be their maximum degree.  Given $T_1$ and $T_2$, a maximum agreement
  subtree of $T_1$ and $T_2$ can be computed in $O(nd^2\log{d}\log^2{n})$
  time or alternatively in $O(nd\sqrt{d}\log^3{n})$ time.  Thus, if $d$ is
  bounded by a constant, a maximum agreement subtree can be computed in
  $O(n\log^2n)$ time.
\end{theorem}
\begin{proof}
  By Theorem~\ref{thm_zero_zero}, the algorithms in
  \S\ref{sec_reductions}--\ref{sec_one_one} compute $\rr(T_1,T_2)$ within
  the desired time bounds.  With straightforward modifications, these
  algorithms can compute a maximum agreement subtree within the same time
  bounds.
\end{proof}

The next lemma establishes a reduction from the longest common subsequence
problem to that of computing a maximum agreement subtree.
\begin{lemma}\label{lem_lower}
  Let $M_1=x_1,\ldots,x_n$ and $M_2=y_1,\ldots,y_n$ be two sequences.
  Assume that the symbols $x_i$ are all distinct and so are the symbols
  $y_j$.  Then, the problem of computing a longest common subsequence of
  $M_1$ and $M_2$ can be reduced in linear time to that of computing a
  maximum agreement subtree of two binary evolutionary trees.
\end{lemma}
\begin{proof}
Given $M_1$ and $M_2$, we construct two binary evolutionary trees $T_1$ and
$T_2$ as follows.  Let $z_1$ and $z_2$ be two distinct symbols different
from all $x_i$ and $y_i$. Next, we construct two paths
$P_1=u_1,\ldots,u_{n+1}$ and $P_2=v_1,\ldots,v_{n+1}$.  $T_1$ is formed by
making $u_1$ the root, attaching $x_i$ to $u_i$ as a leaf, and attaching
$z_1$ and $z_2$ to $u_{n+1}$ as leaves.  Symmetrically, $T_2$ is formed by
making $v_1$ the root, attaching $y_i$ to $v_i$, and attaching $z_1$ and
$z_2$ to $v_{n+1}$.  The lemma follows from the straightforward one-to-one
onto correspondence between the longest common subsequences of $M_1$ and
$M_2$ and the maximum agreement subtrees of $T_1$ and $T_2$.
\end{proof}
We can use Lemma~\ref{lem_lower} to derive lower complexity bounds for
computing a maximum agreement subtree from known bounds for the longest
common subsequence problem in various models of computation
\cite{AHU76,AG87, Goldberg82, Hirschberg75, Hirschberg77, HS77, WC76}.
This paper assumes a comparison model where two labels $x$ and $y$ can be
compared to determine whether $x$ is smaller than $y$ or $x=y$ or $x$ is
greater than $y$.  Since the longest common subsequence problem in
Lemma~\ref{lem_lower} requires $\Omega(n\log{n})$ time in this model
\cite{Hirschberg78}, the same bound holds for the problem of computing a
maximum agreement subtree of two evolutionary trees where $d$ is bounded by
a constant.  It would be significant to close the gap between this lower
bound and the upper bound of $O(n\log^2n)$ stated in
Theorem~\ref{thm_tree}.  Recently, Farach, Przytycka and Thorup
\cite{FPT95} independently developed an algorithm that runs in
$O(n\sqrt{d}\log^3n)$ time. For binary trees, Cole and Hariharan
\cite{CH96} gave an 
${O}(n\log{n})$-time algorithm. It may be possible to
close the gap by incorporating ideas used in those two results and this
paper.

\section*{Acknowledgments}
The author is deeply appreciative for the extremely thorough and useful
suggestions given by the anonymous referee.  The author thanks Joseph
Cheriyan, Harold Gabow, Andrew Goldberg, Dan Gusfield, Dan Hirschberg, Phil
Klein, Phil Long, K.~Subramani, Bob Tarjan, Tandy Warnow for helpful
comments, discussions and references.

\bibliographystyle{siam}
\bibliography{all}

\end{document}